\documentclass[journal]{IEEEtran}

\usepackage{caption}
\usepackage{subcaption}
\usepackage{amsmath,amsfonts}
\usepackage{array}
\usepackage[]{subfig}
\usepackage{textcomp}
\usepackage{stfloats}
\usepackage{url}
\usepackage{verbatim}
\usepackage{graphicx}
\usepackage{cite}
\usepackage{verbatim}
\usepackage{amsmath,amsfonts,amssymb}
\usepackage{xcolor}
\usepackage{stfloats}

\usepackage{multirow}
\usepackage{makecell}
\usepackage{adjustbox}
\usepackage{booktabs}
\usepackage{enumerate}
\usepackage{enumitem}
\usepackage{mathtools}
\usepackage{algpseudocode}
\usepackage{setspace}
\usepackage{comment}
\usepackage{color}
\usepackage{xspace}

\usepackage{flushend}
\usepackage{tabularx}
\usepackage{pifont}
\usepackage{marvosym}
\usepackage{balance}   

\allowdisplaybreaks
\newcommand{\pname}{\texttt{xSlice}\xspace}

\begin{document}
\newcommand{\rev}[1]{\textcolor{black}{#1}}

\markboth{Published in IEEE Transactions on Networking}%
         {Yan \MakeLowercase{\textit{et al.}}: Your Paper Title}

\title{Near-Real-Time Resource Slicing for QoS Optimization  in 5G O-RAN using Deep Reinforcement Learning}

\author{Peihao Yan, 
Jie Lu,
Huacheng Zeng,
and
Y. Thomas Hou\\
\thanks{P.~Yan and H.~Zeng are with the Department of Computer Science and Engineering at Michigan State University, East Lansing, MI 48824.
Y.T. Hou is with the Bradley Department of Electrical and Computer Engineering, Virginia Tech, Blacksburg, VA 24061.
Corresponding author: Huacheng~Zeng (hzeng@msu.edu).}

\thanks{
The work of P.~Yan, J.~Lu and H.~Zeng 
was supported in part by 
NSF under Grant ECCS-2434001 and Grant CNS-2312448 and in part by the National Telecommunications and Information Administration (NTIA) Public Wireless
Supply Chain Innovation Fund (PWSCIF) under Award \# 26-60-IF010.
The work of Y.T.~Hou was supported in part by 
NSF Grants ECCS-2434000 and CNS-2312447 and Virginia Commonwealth Cyber Initiative (CCI).
}
}




\maketitle

\begin{abstract}
Open-Radio Access Network (O-RAN) has become an important paradigm for 5G and beyond radio access networks. This paper presents an xApp called \pname for the Near-Real-Time (Near-RT) RAN Intelligent Controller (RIC) of 5G O-RANs. \pname is an online learning algorithm that adaptively adjusts MAC-layer resource allocation in response to dynamic network states, including time-varying wireless channel conditions, user mobility, traffic fluctuations, and changes in user demand. To address these network dynamics, we first formulate the Quality-of-Service (QoS) optimization problem as a regret minimization problem by quantifying the QoS demands of all traffic sessions through weighting their throughput, latency, and reliability. We then develop a deep reinforcement learning (DRL) framework that utilizes an actor-critic model to combine the advantages of both value-based and policy-based updating methods. A graph convolutional network (GCN) is incorporated as a component of the DRL framework for graph embedding of RAN data, enabling \pname to handle a dynamic number of traffic sessions. We have implemented \pname on an O-RAN testbed with 10 smartphones and conducted extensive experiments to evaluate its performance in realistic scenarios. Experimental results show that \pname can reduce performance regret by 67\% compared to the state-of-the-art solutions.
Source code is available 
at \url{https://github.com/xslice-5G/code}.
\end{abstract}


\section{Introduction}

Resource slicing is a key technique for cellular networks to optimize the allocation of network resources to meet diverse service demands \cite{zhang2017network}. This technique involves partitioning a single physical network into multiple virtual/logic slices, each tailored to specific applications or user demands. By leveraging resource slicing, network operators can efficiently manage and isolate different types of traffic, ensuring high performance and reliability for varied services such as 
Enhanced Mobile Broadband (eMBB),
Ultra-Reliable Low Latency Communication (URLLC), 
massive Machine-Type Communication (mMTC), 
and mission-critical communications. This technique not only enhances network flexibility and scalability but also improves overall efficiency by reducing congestion and prioritizing critical applications. As cellular networks evolve towards 6G and beyond, resource slicing becomes pivotal in supporting the diverse and dynamic needs of emerging ecosystems to enable a more adaptive and resilient network infrastructure.

Cellular networks are inherently dynamic, making online learning approaches particularly appealing for the management of network resources. These approaches can quickly and adaptively adjust resource allocation based on both short-term and long-term traffic and network conditions as well as users' Qualify-of-Service (QoS) demands. Although resource slicing has been explored within the context of Open-Radio Access Network (O-RAN) \cite{garcia2021ran,bonati2021intelligence,niknam2022intelligent,foukas2023taking}, existing approaches to resource slicing are either limited to simulation-based online learning (e.g., \cite{ EDF, yan2019intelligent,  LACO, Channel-Aware}) or offline training using realistic O-RAN datasets (e.g., \cite{foukas2017orion, polese2022colo}). 
Design and validation of online-learning-based resource slicing solutions remain limited for real-world 5G RANs
(see Table~\ref{tab:relwork}).
This scarcity of online learning methods in realistic O-RAN systems underscores the underlying challenges and highlights the need to fill this gap for O-RAN. Such innovations must effectively meet users' QoS demands while adapting to network dynamics, including traffic volatility and user mobility.

In this paper, we introduce \pname, a resource slicing xApp designed for the Near-Real-Time (Near-RT) RAN Intelligent Controller (RIC) within an O-RAN system. 
\pname operates as an online algorithm within the Near-RT RIC, managing resource slicing at the Media Access Control (MAC) layer by monitoring performance metrics of the RAN. It functions as an independent module on the O-Cloud platform, interfacing with the RAN through standard O-RAN interfaces (e.g., E2), ensuring full compatibility with the O-RAN architecture.
Leveraging the feedback loop of the Near-RT RIC with a response time ranging from ten milliseconds to one second, \pname can adaptively partition network resources to optimize QoS for users with diverse QoS demands, while effectively addressing network dynamics such as traffic volatility and user mobility.


%
Design of \pname presents two main challenges.
The first challenge lies in the time-varying nature of wireless channel conditions and the diverse QoS demands of different traffic sessions (applications). 
Wireless connections are notorious for time-varying path loss and unpredictable fading. Even if a user remains stationary, their wireless channel can fluctuate over time due to changes and movement in their surroundings, as well as temperature variations affecting their RF devices. Additionally, different users have varying QoS demands in terms of connection throughput, latency, and reliability. The combination of time-varying channel conditions and diverse QoS demands presents a complex challenge for network optimization through resource slicing.

\begin{table*}
\renewcommand{\arraystretch}{1.15}
\scriptsize
\color{black}
\centering
\caption{\rev{
Representative work on 5G network slicing in the literature.}}
\resizebox{\linewidth}{!}{
\begin{tabular}{|c@{\hspace{2pt}}|l@{\hspace{2pt}}|l@{\hspace{2pt}}|l@{\hspace{2pt}}|c@{\hspace{4pt}}|c@{\hspace{3pt}}|c@{\hspace{4.5pt}}|}
\hline
\textbf{Reference} & \textbf{Objective} & \textbf{Key idea} &  \textbf{Evaluation testbed} & \textbf{\!\!\!\!OTA?\!\!\!\!} & \textbf{\!\!\!UE device\!\!\!} & \textbf{\!\!\!\!\# of UEs\!\!\!\!}   \\ \hline

NVS\cite{NVS} & 5G Slice isolation and utilization & Calculate a reserved cumulative bandwidth & PicoChip\cite{picochip}& Yes & Beceem chip & 5 \\ \hline

EDF\cite{EDF} & Slice scheduling & Earliest deadline first &Simulation& No   & N/A & N/A  \\ \hline

IRSS\cite{yan2019intelligent} & Resource utilization & LSTM + Q-learning tackle network dynamics & Simulation & No &N/A & N/A \\ \hline   

Orion\cite{foukas2017orion} & Design base station hypervisor & Introduce a abstractions for virtualization &  FlexRAN platform\cite{flexran} &Yes & Smartphones & 4 \\ \hline  

LACO\cite{LACO} & Guarantees latency and throughput & Multi-armed-bandit &  srsLTE\cite{gomez2016srslte} \& simulation & No &Tablet & N/A  \\ \hline

CoIO-RAN\cite{polese2022colo} & Train and test ML solution xApps & Collecting datasets at scale + DRL &  Arena platform \cite{bertizzolo2020arena} &Yes & Smartphones & 3 \\ \hline 

RadioSaber\cite{Channel-Aware} & Inter-slice and enterprise slicing  &Channel-aware decision making & Open5gs\cite{mamushiane2023deploying}\&Simulation & No & N/A  & N/A  \\ \hline

Zipper\cite{app_based} & Bandwidth efficiency & RNN predict SNR + DNN map states to PRBs & FlexRAN\cite{foukas2016flexran}& Yes  & N/A & N/A \\ \hline

IQRA \cite{mhatre2024intelligent} & Resource allocation in network slicing & limited-action-space DQN & Simulation  & No & N/A & N/A \\ \hline

PW-DRL \cite{cai2023deep} & Slice resource allocation & Per-slice DRL for dynamic allocation & Simulation & No & N/A  & N/A \\ \hline


GNN+TD3 \cite{liu2024achieving} & Energy-efficient 5G slicing & Resource allocation with slicing-aware optimization & Simulation & No & N/A & N/A \\ \hline

\textbf{\pname (Ours)}& 5G resource slicing & GCN extract sessions features + PPO learn policy & OpenAirInterface\cite{openairinterface5g}  & Yes& Smartphones & 10 \\  \hline
\end{tabular}
}
\label{tab:relwork} 
\end{table*}

To address this challenge, we formulate a regret-based optimization objective function that characterizes the QoS demands of a traffic session by incorporating weights for its throughput, latency, and reliability demands. This approach enables network operators to set the priority of each application by adjusting its QoS weights. Based on this regret-based objective, we design a Deep Reinforcement Learning (DRL) framework for online resource slicing optimization.
The core of the DRL framework is a xApp within the Near-RT RIC, which makes resource slicing decisions based on its observations of Key Performance Metrics (KPMs) provided by the RAN. To enhance the learning stability and efficiency, we adopt an actor-critic architecture for the DRL framework. 
This actor-critic architecture combines the advantages of both value-based and policy-based methods, making it possible for the DRL framework to rapidly adjust its resource slicing decisions based on dynamic wireless channels and diverse QoS demands. 


The second challenge lies in network dynamics, particularly the variability of active users and their traffic sessions. In a cellular network, both the number of active users and the characteristics of their traffic sessions fluctuate over time, making it impossible for the DRL network to take the raw per-session state data from the RAN as input for exploration and exploitation. 
To address this challenge, we propose using a Graph Convolutional Network (GCN) approach to model the relationships among the MAC-layer data and the demand of traffic sessions, with each node in the graph representing a traffic session. The GCN captures and learns the underlying patterns in the MAC data by aggregating and transforming node features based on their neighbors. This results in powerful node embeddings, allowing the MAC data of all traffic sessions to be represented using a unified yet compact hidden layer, which serves as the input for the DRL model. 
By integrating the GCN with the actor-critic DRL network, our method efficiently processes MAC\&KPM data from any number of traffic sessions, enabling effective online resource slicing optimization.

We have implemented \pname as an xApp in a realistic 5G O-RAN network where a gNB serves ten Commercial Off-The-Shelf (COTS) smartphones. 
\pname communicates with the RAN via an E2 interface to obtain the MAC\&KPM data including throughput, latency, block error rate (BLER), etc., which are used as the input to make resource slicing decisions. %
Extensive Over-The-Air (OTA) experimental results show that \pname consistently outperforms existing solutions.
On average, it reduces the performance regret by 67\% compared to the state-of-the-art solutions.

The main contributions of this paper are as follows.

\begin{itemize}[leftmargin=0.2in,itemsep=0in,topsep=0.0in]

\item
\rev{We design and implement a novel DRL-based framework that jointly addresses adaptability and multi-session scalability for resource slicing in 5G O-RAN. Unlike prior DRL approaches that are either based on simulation evaluation, our framework operates in a live O-RAN system using real KPM data.}

\item
\rev{We introduce a GCN-based adapter module for the DRL framework. This adapter enables the KPM data and demand from a dynamic number of traffic sessions to be represented using a unified yet compact hidden layer, enhancing the efficiency of \pname in handling a dynamic number of traffic sessions.}


\item
\rev{We implement \pname as a Near-RT RIC xApp in a 5G O-RAN testbed and conduct extensive over-the-air experiments with smartphones running realistic application traffic. Results show that \pname enables fast, Near-RT decision-making with strong adaptability, efficiency, and robustness under diverse and time-varying network conditions.}

\end{itemize}

\section{Related Work}

RAN slicing has been extensively studied in the context of 5G to improve network flexibility and optimize performance.
In general, existing research can be categorized into three main categories: 
spectrum slicing \cite{hong2012picasso,foukas2017orion,NVS,Channel-Aware,app_based,TLRLO-RANSlicing}, 
infrastructure slicing \cite{jiang2022probabilistic, bega2017optimising, wu2022survey}, 
and network resource slicing \cite{foukas2017network, ebrahimi2024resource}
Our literature survey focuses on network resource slicing.
\rev{Table~\ref{tab:relwork} summarizes the differences between \pname and the most closely related prior work.}

\noindent\textbf{RAN Slicing.}
A variety of network resource slicing approaches have been proposed, highlighting their potential for efficient resource management in RANs.
Orion \cite{foukas2017orion} introduced a cost-effective network slicing framework for multi-service RANs, enabling adaptive resource provisioning based on slice-specific demands.
Network Virtualization Substrate (NVS) \cite{NVS,kokku2010nvs} is a widely adopted slicing technique that partitions resources by assigning maximum weights to each slice in every time slot. These weights reflect the marginal utility of a slice relative to the cumulative fraction of resources it has received.
RadioSaber \cite{Channel-Aware} proposes a channel-aware inter-slice resource scheduler that exploits the variability of wireless channel conditions across frequency bands, users, and time. This method relies on prior knowledge of channel quality per slice.
Zipper \cite{app_based} is a recent RAN slicing strategy that focuses on application-level resource allocation instead of user-specific allocation. It tracks the dynamic network state for each user and employs an efficient algorithm to allocate bandwidth across slices to meet individual application requirements.
Unlike these prior works, \pname is an online learning framework that enables adaptive decision making for resource slicing.

\noindent\textbf{DRL for 5G Network Slicing.}
DRL has been increasingly explored for network slicing in RANs, with progressively more sophisticated approaches \cite{TLRLO-RANSlicing, FLforRA2024, xappGNNRL2021, IEEEcomDLO-RANSlicing}.
IRSS \cite{yan2019intelligent} introduced a learning-based resource scheduling mechanism that predicts and allocates resources based on current and anticipated demand.
The work in \cite{globecom2022DLresourceslicing} proposed a joint global model that is decomposed into local Q-tables for different xApps, enabling decentralized resource allocation decisions.
LACO \cite{LACO} leverages the exploration–exploitation trade-off using a multi-armed bandit (MAB) orchestrator to make adaptive resource slicing decisions without requiring prior knowledge of traffic demand or channel conditions.
\rev{\cite{cai2023deep} proposed a prediction-aided weighted DRL (PW-DRL) framework for online, dynamic RAN slicing, jointly optimizing power allocation and user admission based on varying service priorities. \cite{mhatre2024intelligent} developed a DQN-based, QoS-aware intra-slice allocation strategy with user association in O-RAN, improving eMBB throughput and reducing URLLC latency.
However, all of the above approaches remain theoretical, as they do not adopt practical input/output datasets for DRL. Moreover, they were validated solely through simulations rather than real-world experimentation, leaving their practical performance unverified.}



\noindent
\rev{
\textbf{GNN-DRL in O-RAN.}
Several pioneering studies have integrated Graph Neural Networks (GNNs) with DRL for online network optimization in O-RAN.
For instance, \cite{tam2024graph} proposed a GNN-assisted DRL framework for managing collaborative learning workloads in virtualized O-RAN environments.
\cite{liu2024achieving} introduced a GNN-DRL approach to enhance energy efficiency by predicting workloads and managing base station sleep modes.
\cite{orhan2021connection} explored the use of GNN-DRL for connection management via xApps in the RIC.
These works, however, pursue different optimization objectives compared to \pname.}

\noindent\textbf{xApps for O-RAN Resource Allocation.}  
There is a substantial body of literature on MAC-layer resource allocation for cellular networks. Recently, efforts have shifted toward O-RAN systems \cite{polese2023understanding, FlexAppChen}, particularly through the implementation of xApps in Near-RT RIC \cite{ntassah2023xapp, thieu2023demystifying}. AI-based xApps have been developed to optimize various network metrics, including user scheduling \cite{compare_performance, lvproviding}, resource allocation \cite{BayesianRadio,BayesianMulti-ArmedRadiorResource, BayesianForVRAN, RLresourceallocation, globecom2022DLresourceslicing, sciancalepore2017mobile, hou2023efficient}, spectrum utilization \cite{priority-based}, and traffic prediction \cite{proportionally-fair, spink1996multiple}. 
However, to date, most existing learning-based xApps rely on offline training, and little progress has been made on online training for xApps in Near-RT RIC. \pname addresses this gap.
CoIO-RAN \cite{polese2022colo} is closely related to \pname. It proposes a DRL-based framework for resource allocation for each slice. However, CoIO-RAN primarily aims to provide a programmable platform using xApps for generating large-scale datasets and evaluating the performance of machine learning models. It utilizes a conventional DRL model without a GCN module. 
\rev{In contrast, \pname focuses on optimizing a defined regret objective through resource slicing using a Near-RT xApp.}

\section{Preliminaries}
In this section, we first offer a primer on O-RAN and then elaborate on the need for adaptive resource slicing operations in O-RANs. 

\subsection{A Primer on O-RAN}

\noindent
\textbf{O-RAN.}
O-RAN is a transformative architecture for building and managing radio access networks with open, interoperable standards, fostering innovation and reducing costs by allowing multi-vendor solutions. It has potential to enhance network flexibility and efficiency by enabling more collaborative and modular network architectures.
A typical 5G O-RAN deployment consists of three main components: Radio Unit (RU), Distributed Unit (DU), and Centralized Unit (CU). The RU is usually implemented with specialized hardware, while the CU and DU functions \cite{xing2023enabling} are implemented as software applications running on cloud servers. Each RU serves a single cell and connects to the DU via a front-haul network\cite{budhdev2021fsa,bonati2021scope,lazarev2023resilient}. A gNodeB (gNB) can interface with multiple CUs, DUs, and RUs. 
Additionally, the Near-RT RIC, which can be physically separated from the gNB and operate within a cloud server, enhances the programmability of the RAN. It is for near-real-time control and management of the RAN, utilizing E2 interfaces and xApps it hosts.

\noindent\textbf{SDU buffer.}
Service Data Units (SDUs) represent the data packets exchanged between the RLC and MAC layers in RAN.  
Within the RLC layer, SDUs are segmented into smaller units for transmission and reassembled upon reception. The MAC layer is responsible for scheduling and multiplexing these SDUs onto the physical channel, effectively coordinating their transmission based on the received SDUs from the RLC layer.
The calculation of KPM is based on the data in the SDU buffer.


\noindent\textbf{Understand SNR and CQI.}\label{subsec:CQI} 
For 5G gNB, Channel State Information (CSI) reports are crucial for identifying changes in UE state. However, CSI and Signal-to-Noise Ratio (SNR) can only be obtained for the uplink channels, such as Physical Uplink Control Channel (PUCCH) and Physical Uplink Shared Channel (PUSCH), but cannot be obtained for the downlink channels. 
The Channel Quality Indicator (CQI) measures the downlink channel quality and is reported by the UE to the gNB.
Specifically, the UE calculates the CQI index based on the estimated downlink CSI, accounting for factors such as SNR, interference levels, and channel conditions. 
 
\subsection{Need for Resource Slicing}

\noindent\textbf{Diverse QoS Demands.}
5G is designed to support a wide range of traffic sessions, each with its own unique QoS demands in terms of throughput, latency, and reliability. Based on their QoS demands, traffic sessions can be classified into different categories:
(i) eMBB: High data rates for applications such as video streaming and virtual reality.
(ii) URLLC: Extremely low latency and high reliability for applications like autonomous vehicles and industrial automation.
(iii) mMTC: High device density and low power consumption for IoT devices and sensors.

\noindent\textbf{Service Level Agreements (SLA).}
SLAs define the expected level of service between RAN and applications. One of the most significant advancements with 5G is its capacity to cater to the diverse SLA demands through resource slicing. 
With the proliferation of 5G networks, the assurance of performance metrics for SLAs has become increasingly critical, particularly in terms of latency, jitter, and bandwidth. Each session is associated with specific demands that must be met.






\section{Problem Formulation}









\subsection{Regret-based Formulation}
The objective of this work is to design an efficient resource slicing xApp for the Near-RT RIC that can adapt resource partitioning to the dynamic QoS demands of UEs in an O-RAN system. 
Denote $\mathcal{S}$ as the set of traffic sessions within the gNB to support the applications of all the active UEs, 
with $S = |\mathcal{S}|$. 
One UE may maintain one or multiple traffic sessions with different QoS demands. 
For instance, a UE may have online gaming and file downloading applications at the same time, where online gaming belongs to URLLC traffic while file downloading belongs to eMBB traffic. 

To better support the traffic sessions based on their QoS demands, the O-RAN partitions its time-frequency resource into $K$ slices, with each of its traffic sessions being assigned to one slice. 
Denote $\mathcal{K}$ as the set of slices in the O-RAN, with $K = |\mathcal{K}|$. 
Denote $\mathcal{S}_k$ as the set of sessions assigned to slice $k \in \mathcal{K}$. 
For the sessions in slice $k\in \mathcal{K}$, denote $P_k$ as their throughput demand, $T_k$ as their delay demand, and $Z_k$ as their reliability (BLER) demand.

To make online decisions on resource slicing, the xApp needs to query the RAN (via the E2 interface) for KPM acquisition through query iterations. Let \(T\) denote the time duration of one query round and \( t \) denote xApp's query iteration index. For session \( i \in \mathcal{S} \), let \( \rho_i(t) \) denote its achieved throughput, \( \tau_i(t) \) denote its experienced time delay, and \( \zeta_i(t) \) denote the BLER of its transmission, all within query round \( t \).
All these KPMs are computed in RAN and reported to xApp in response to its queries.

We now model the regret of session $i$ in slice $k$ and time slot $t$, which includes three components: throughput, delay, and reliability (BLER).
For throughput, if its achieved throughput is greater than its demand, the regret is zero; otherwise, the regret is defined as the normalized throughput deficit (i.e., $\frac{P_k - \rho_i(t)}{P_k}$). 
Combining these two cases, its throughput regret is modeled as: 
$\max\big(\frac{P_k - \rho_i(t)}{P_k}, 0\big)$. 
Similarly, the delay and reliability regrets are modeled as:
$\max\big(\frac{\tau_i(t) - T_k}{T_k}, 0\big)$ and 
$\max\big(\frac{\zeta_i(t) - Z_k}{Z_k}, 0\big)$, 
respectively.

Let $r_{k}^{\mathrm{[p]}}(t)$ denote the throughput regret of the sessions in slice $k$ and time slot $t$,
$r_{k}^{\mathrm{[d]}}(t)$ denote their delay regret,
and
$r_{k}^{\mathrm{[r]}}(t)$ denote their reliability regret.
Then, we have 
\begin{align}
& r_{k}^{\mathrm{[p]}}(t) = \sum_{i \in \mathcal{S}_k} \max\big(\frac{P_k - \rho_i(t)}{P_k}, 0\big), \nonumber \\
& r_{k}^{\mathrm{[d]}}(t) = \sum_{i \in \mathcal{S}_k} \max\big(\frac{\tau_i(t)-T_k}{T_k}, 0\big), \nonumber \\
& r_{k}^{\mathrm{[r]}}(t) = \sum_{i \in \mathcal{S}_k} \max\big(\frac{\zeta_i(t)-Z_k}{Z_k}, 0\big).  \nonumber 
\end{align}


To model the different QoS demands of different slices, we introduce a non-negative vector to denote the weights of their throughput, delay, and reliability regrets. 
Specifically, let $\lambda_{k}^{\mathrm{[p]}}$, $\lambda_{k}^{\mathrm{[d]}}$, $\lambda_{k}^{\mathrm{[r]}}$ denote
the throughput, delay, and reliability weights of the sessions in slice $k$.
Network operators can use these weights to adjust the priorities of throughput, delay, and reliability in online optimization.
Incorporating the weights, the total regret of all sessions in the O-RAN can be written as:
\begin{equation}
\!\!r(t) 
\!=\! 
\sum_{k \in \mathcal{K}} \sum_{i \in \mathcal{S}_k}
\left(
\lambda_{k}^{\mathrm{[p]}} r_{k}^{\mathrm{[p]}}(t)
+
\lambda_{k}^{\mathrm{[d]}} r_{k}^{\mathrm{[d]}}(t)
+
\lambda_{k}^{\mathrm{[r]}} r_{k}^{\mathrm{[r]}}(t)
\right) 
\,.
\label{eq:regret_all}
\end{equation}

Consider the case where the O-RAN has light traffic. 
In this case, the QoS demand of all traffic sessions can be met. 
As a result, the regret in Equation \eqref{eq:regret_all} will be zero. 
To accommodate such light-traffic cases, we introduce an auxiliary objective of minimizing the number of required physical resource blocks (PRBs), so that the unused PRBs can be maximized to admit new UEs. 
Specifically, let $n_k(t)$ denote the number of PRBs used in slice $k$.
Then, we define the following utilization function as the auxiliary function:
\begin{equation}
u(t) 
= \sum_{k \in \mathcal{K}} \frac{1}{n_k(t) + C}
\,,
\label{eq:utilization_all}
\end{equation}
where $C$ is a non-negative constant that is empirically set.


Combining the regret function in Equation~\eqref{eq:regret_all} and the utilization function in Equation~\eqref{eq:utilization_all}, we formulate this optimization problem as a regret minimization problem:
\begin{subequations}
\begin{align}
    \min_{n_k(t)} ~~~& r(t) - \; u(t)  \\
    \mbox{s.t.} ~~~& \sum_{k \in \mathcal{K}} n_k(t) \le N^\mathrm{[rb]},
\end{align}
\label{eq:opt1}
\end{subequations}
\!\!where $N^\mathrm{[rb]}$ is the total number of PRBs in O-RAN. 
In this optimization problem, we aim to find the optimal decision on $n_k(t)$ at the end of iteration $(t\!-\!1)$. 
$n_k(t')$, $1 \le t' < t$, is the actions that have been taken in previous iterations.
$\rho_i(t')$, $\tau_i(t')$ and $\zeta_i(t')$, $1 \le t' < t$,  are the observations corresponding to previous actions.
$n_k(t)$ is the optimization variables (decisions to make in the current iteration).
The values of $\rho_i(t)$, $\tau_i(t)$ and $\zeta_i(t)$ are dependent upon $n_k(t)$ but their relationship is complex and unknown.
$C$, $\lambda_k^\mathrm{[p]}$, $\lambda_k^\mathrm{[d]}$, $\lambda_k^\mathrm{[r]}$ are constants.

\vspace{-0.1in}
\subsection{Challenges}

\begin{figure}
    \centering
    \begin{subfigure}[b]{1.6in}
        \centering
        \includegraphics[width=1.7in, trim= 0 0 0 0, clip]{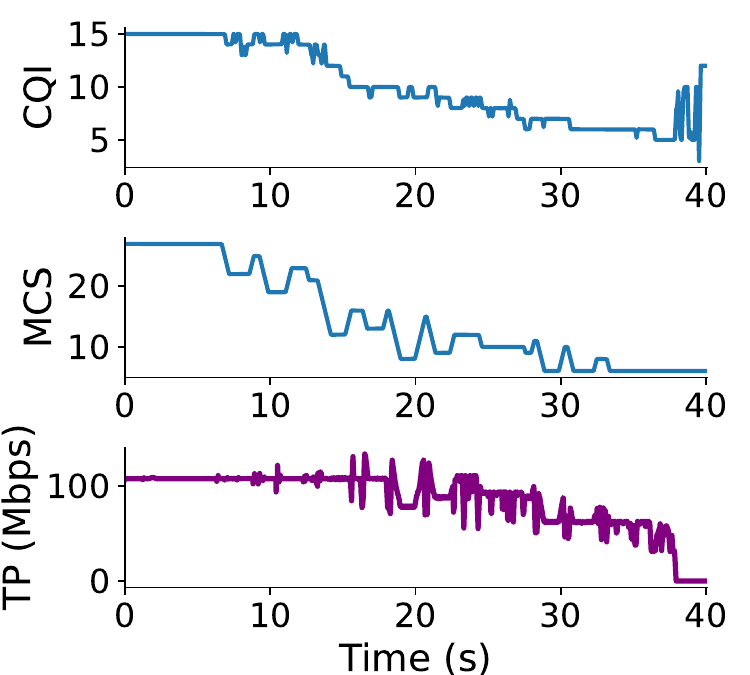}
        \caption{UE moves farther away from gNB.}
        \label{fig:throughputdecrease}
    \end{subfigure}
    \hfill
    \begin{subfigure}[b]{1.6in}
         \centering
         \includegraphics[width=1.7in]{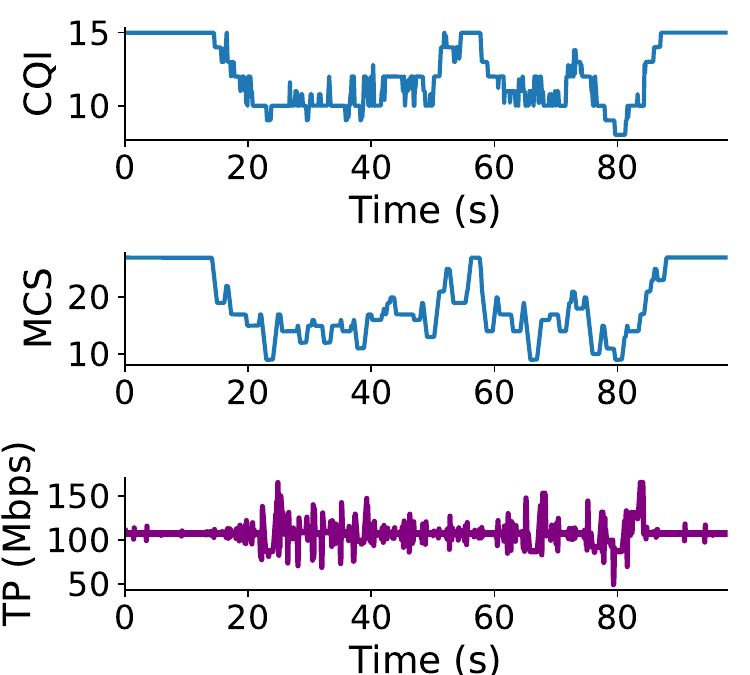}
        \caption{UE is stationary but people are walking in its proximity.}
         \label{fig:throughputnostable.}
    \end{subfigure}
    \caption{Downlink CQI, MCS, and throughput (TP) changes of an UE in two cases.}
    \label{fig:throughtputchange}
\end{figure}

\noindent
\textbf{Time-varying Wireless Channel Conditions.}
User mobility is a key feature of O-RAN systems, and the condition of wireless channels is time-varying, depending upon their path loss, shadow fading, and other factors. 
To understand the channel dynamics, we conducted experiments using a Samsung smartphone and a gNB to measure the channel quality and the achievable throughput of UE in two cases: 
(i) UE is moving farther away from gNB;
and
(ii) UE is stationary but people are walking in its proximity.
Fig.~\ref{fig:throughtputchange} presents the downlink CQI, MCS, and throughput changes of a UE in the two cases.  
It is evident that the channel quality and achievable throughput of the UE are changing over time in both cases (for both mobile UE and stationary UE). 
Adapting the resource slicing to the change of wireless channel conditions is nontrivial.

\begin{figure}
    \centering 
    \begin{subfigure}[b]{1.6in}
        \centering
        \includegraphics[width=1.6in]{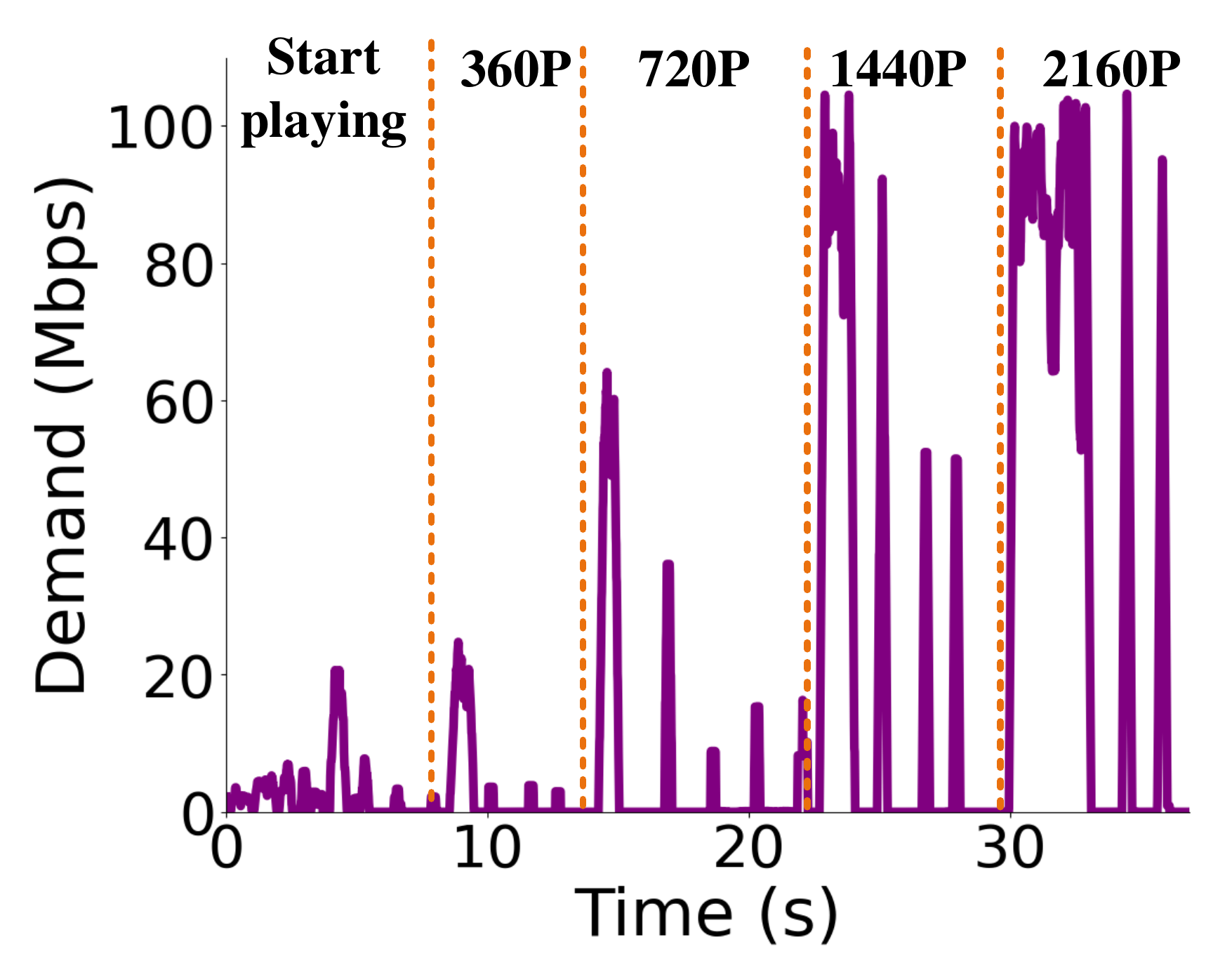}
        \caption{Throughput demand of a YouTube session when a user chooses different video quality.}
        \label{fig:visioyoutube.}
    \end{subfigure}
    \hfill
    \begin{subfigure}[b]{1.6in}
         \centering
         \includegraphics[width=1.7in]{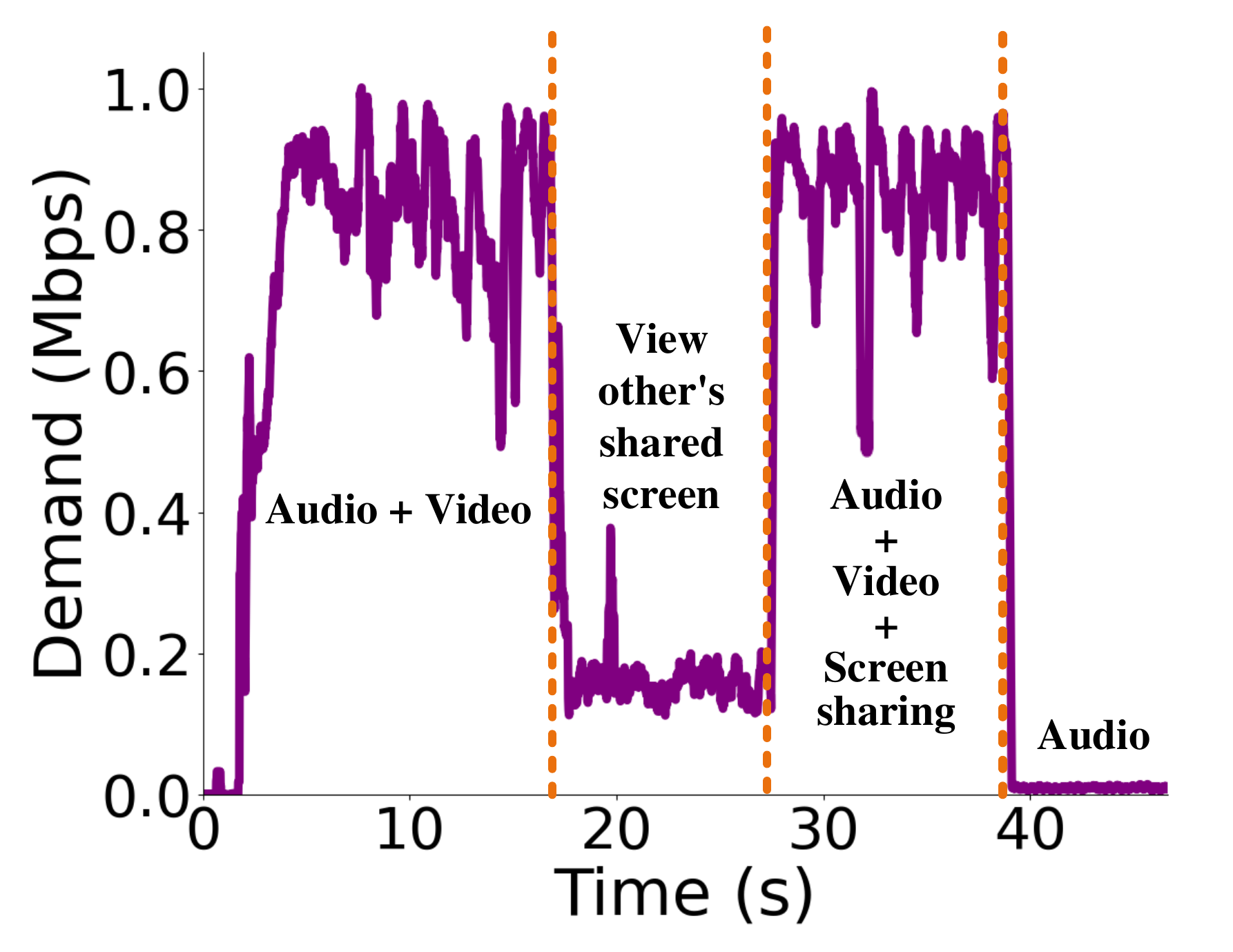}
         \caption{Throughput demand of a Zoom session when a user uses different features.}
         \label{fig:visiozoom.}
    \end{subfigure}
    \caption{Change in throughput demand of YouTube and Zoom traffic sessions over time.}
    \label{fig:demand_change} 
\end{figure}

\noindent
\textbf{Dynamics of QoS Demand.}
The QoS demand of existing traffic sessions may change over time. For instance, a user may change the quality of his/her YouTube video streaming from 360p to 1440p quality, significantly increasing the required throughput of his/her wireless connection. 
To illustrate the QoS demand dynamics of a traffic session (application), we conducted experiments on a smartphone to measure the throughput demand of a YouTube video streaming session and a Zoom teleconferencing session.
Fig.~\ref{fig:demand_change} shows their throughput demand when a user uses different features of the applications. 
It can be seen that the throughput demand of a traffic session is dynamic over time, depending upon the user's operation and activity. 
An efficient resource slicing algorithm must adapt to the QoS demand change of all traffic sessions in terms of throughput, delay, and reliability.

\noindent
\textbf{Newly-Arrival and Terminated Traffic Sessions.}
The data traffic sessions in an O-RAN system are dynamic, influenced by their life cycles, network admission policies, UE activities, etc. 
An existing traffic session may be ended or terminated by its user or server, unexpectedly or normally.
A new traffic session may start when a new connection request is made by a user.

The network dynamics arising from time-varying wireless channel conditions, time-varying QoS demands of individual traffic sessions, random arrivals of new traffic sessions, and unexpected terminations of existing sessions necessitate an efficient online resource slicing solution. This solution must adapt resource allocation to the dynamic network environment by learning its underlying features.

\section{\pname: Design}
In this section, we present \pname to solve the QoS optimization problem in \eqref{eq:opt1}.


\begin{figure*}[ht]
    \centering
    \includegraphics[width=7.0in, trim= 10 3 10 3, clip]{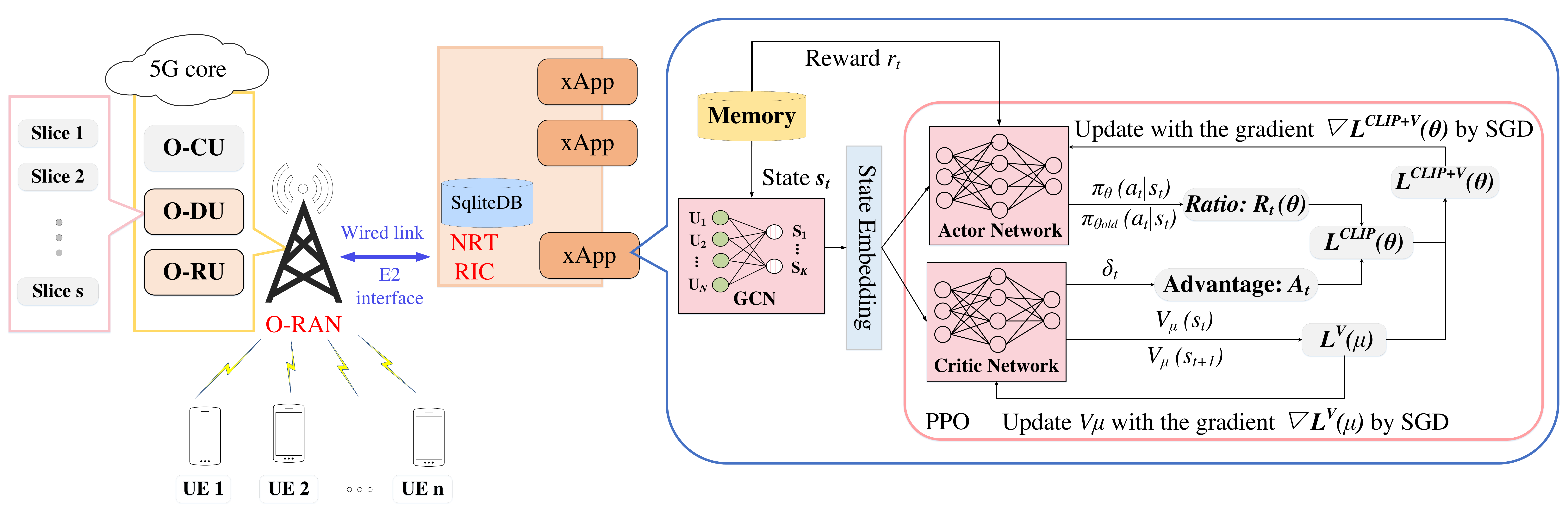}
    \caption{The overview of our proposed DRL framework for resource slicing in an O-RAN system.}
    \label{fig:main_frame}
\end{figure*}

\subsection{Overview}

\noindent\textbf{Input Data.}
Fig.~\ref{fig:main_frame} presents the system architecture of \pname in an O-RAN.
\pname is designed as an xApp running in the Near-RT RIC, communicating with gNB (O-CU, O-DU, and O-RU) via the well-established E2 interface and making resource slicing decisions for O-DU.
\rev{The KPM and MAC data of each UE were obtained from the gNB via the E2 interface in every iteration.}
Table~\ref{tab:kpmmacdata} details the KPM and MAC data that \pname obtains for each individual UE. 
This data is used as the input of \pname to update its resource slicing decisions with the aim of minimizing the regret defined in Equation~\eqref{eq:opt1}.

\begin{table}
\renewcommand{\arraystretch}{1.15}
\caption{The list of KPM and MAC data that \pname obtains for each traffic session.}
\resizebox{\linewidth}{!}{
\begin{tabular}{|c|l|l|l|}
\hline
\multicolumn{1}{|l|}{} & Data & \!\!UL or DL?\!\!\! & Explanation \\ \hline
\!\!\!\multirow{3}{*}{\begin{tabular}[c]{@{}c@{}}KPM \\ data\end{tabular}}\!\!\! & Per-UE throughput\!\!\! & DL & Average data rate achieved by UE in 10 ms.  \\ \cline{2-4} 
 & Per-UE delay & DL & Delay of SDU after being requested. \\ \cline{2-4} 
 & Per-UE PRBs & DL & \# of PRBs assigned to each UE in 10 ms.\\ \hline
\!\!\!\multirow{7}{*}{\begin{tabular}[c]{@{}c@{}}MAC \\ data\end{tabular}}\!\!\! & PUSCH SNR & UL & Quality of signal transmitted by UE. \\ \cline{2-4} 
 & PHR & DL\&UL &  Difference of max Tx power and current usage. \\ \cline{2-4} 
 & MCS & DL\&UL &  modulation index and coding rate for data. \\ \cline{2-4} 
 & BLER & DL\&UL & Percentage of blocks received with errors. \\ \cline{2-4} 
 & Current TBs & DL&  \# of Transport Blocks being Tx-ed in a given time.\!\!\!\! \\ \cline{2-4} 
 & Scheduled RBs & DL& \# of PRBs scheduled for transmission. \\ \hline  
\end{tabular}
}
\label{tab:kpmmacdata}
\end{table}


\noindent\textbf{Bandwidth Parts (BWP).}
The resource slicing decisions from \pname, representing the number of consecutive physical resource blocks (PRBs) for each slice, must be taken by the O-DU to make it effective. 
To ensure compatibility with O-RAN and 5G New Radio (NR), we use the BWP interface for the resource slicing configuration at the O-DU. 
BWP is a feature in 5G NR for the dynamic allocation and management of frequency resources within a given carrier. A BWP represents a contiguous block of spectrum within a carrier, allowing the network to tailor its bandwidth allocation to specific needs and conditions. 
The resource slicing decisions from \pname are then translated to the BWP parameters (e.g., the start index of PRB and the length of PRB for each slice), which are taken by the O-DU to partition the frequency bandwidth.

\begin{figure}[t]
    \centering
    \includegraphics[width=0.9\linewidth]{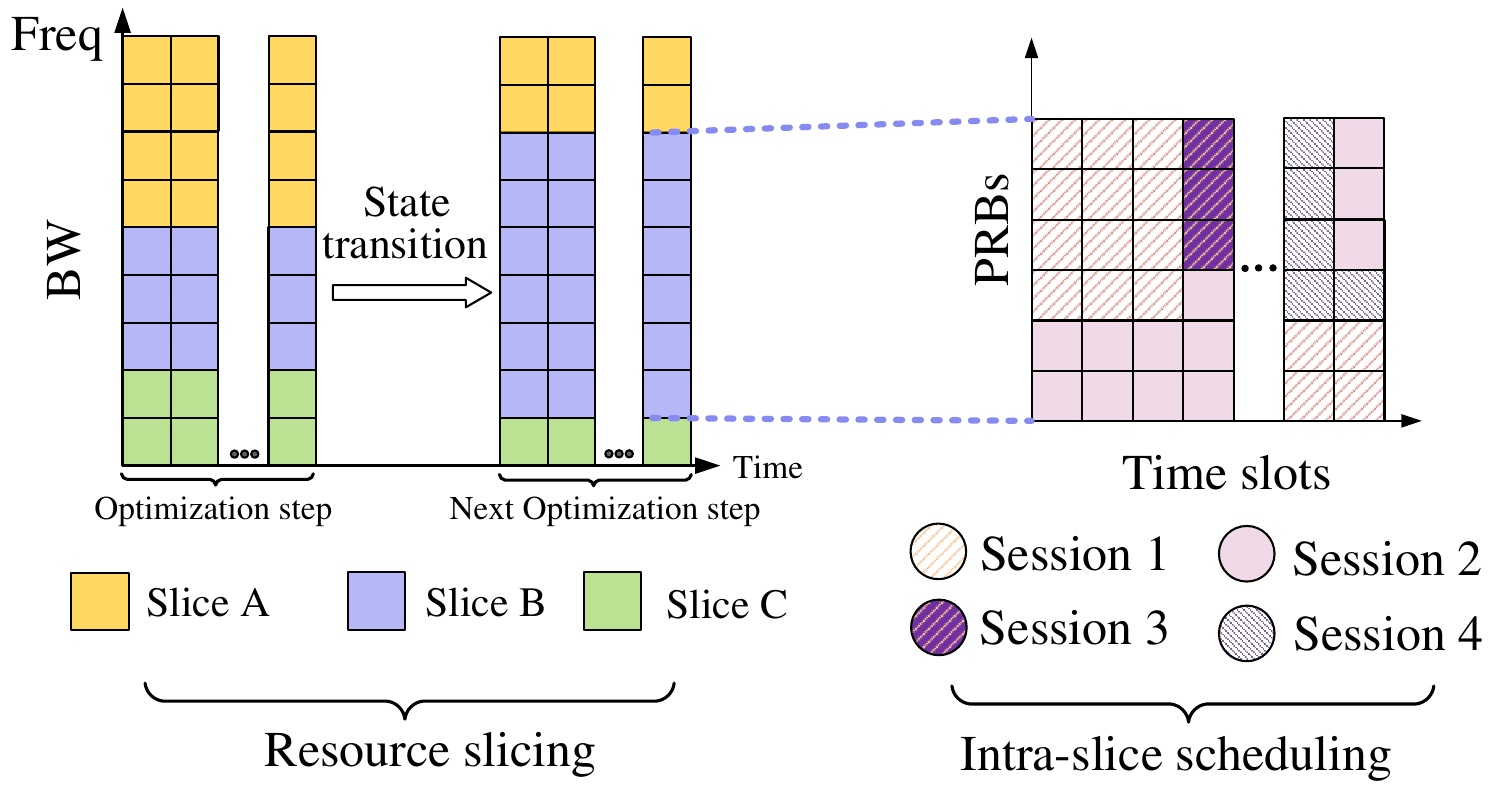}
    \caption{Near-real-time resource slicing \textit{versus} real-time resource scheduling for individual slices.} 
    \label{fig:bwps}
\end{figure}

Fig.~\ref{fig:bwps} illustrates the resource allocation within an O-DU, which involves two time scales: 
\textit{resource slicing} (near-real-time, 10ms to 1s) and \textit{resource scheduling} (real-time, $\le10$ms).
Resource slicing is carried out by \pname and implemented through the BWP interface, while the resource scheduling is managed by O-DU itself for individual slices. 
O-DU may use scheduling algorithms such as
proportional fair,
round robin, 
maximum throughput,
and
earliest deadline first (EDF), 
depending upon the QoS requirements of the traffic sessions in a slice. 

\subsection{GCN Module}
\label{subsec:GCN}

\noindent
\textbf{Motivation.}
Table~\ref{tab:kpmmacdata} lists the input data for \pname in each iteration, which has the following two features:
\textit{(i) The input data is inherently correlated.}  
As shown in Table~\ref{tab:kpmmacdata}, KPM and MAC data such as throughput, delay, BLER, and SNR are tightly coupled. In certain scenarios, one metric can be inferred from others. Therefore, the input KPM and MAC data can be transformed into a compact domain for key feature extraction before being processed by the DRL model.
\textit{(ii) The input data is highly dynamic.}  
This dynamism is evident on two time scales. On a large time scale, the number of active UEs changes over time due to traffic life cycles, user activities, and admission control. On a small time scale, the input data is influenced by the rapidly changing wireless channel conditions of individual UEs.

To accommodate these features, we employ a Graph Convolutional Network (GCN) as the preprocessing step for the DRL model. GCNs\cite{wang2020gcn, Teal2023sigcom, networkplanning2021sigcom} excel at processing graph-structured data by performing convolution operations on nodes and edges, capturing both local and global structures. They aggregate information from neighboring nodes to model dependencies effectively and propagate information throughout the graph. The GCN in \pname will standardize the input data dimension and format by mapping KPM and MAC data into a compact domain, while extracting the key features for the subsequent DRL modules.

\begin{figure} [t]
    \centering
    \includegraphics[width=\linewidth]{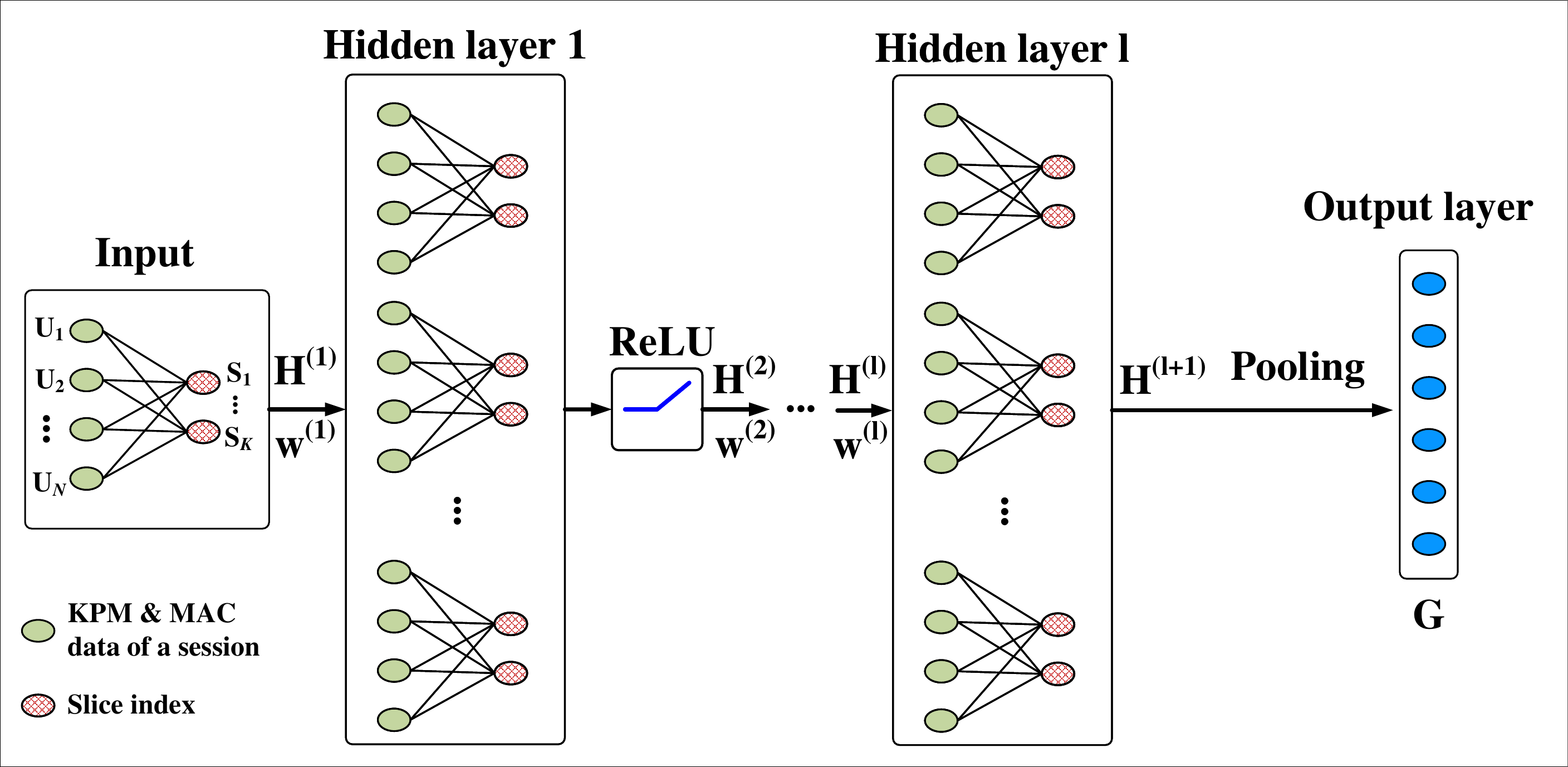}
    \caption{GCN architecture for \pname.}
    \label{fig:gcn}
\end{figure}

\noindent\textbf{Graph Encoding.}
We use a bipartite graph to encode the input data for the GCN, as shown in Fig.~\ref{fig:gcn}. The nodes are divided into two sets: session nodes and slice nodes. 
Each session node represents an existing or dummy session. 
An existing session is encoded with its KPM and MAC data as shown in Table~\ref{tab:kpmmacdata} as its attributes, while a dummy session does not have attributes. 
The edge between a session node and a slice node indicates that the session is assigned to that slice.
This bipartite GCN explicitly exploits the bipartite nature of the graph, focusing on interactions between nodes in these two sets. This allows for efficient processing and representation of the relationships between sessions and slices.



\noindent\textbf{GCN Processing.}
The objective of GCN is to obtain embeddings for the bipartite graph of sessions and slices, characterizing each session's long-term and short-term states in each slice.
Fig.~\ref{fig:gcn} shows the GCN architecture.
The fundamental concept of GCN is to exchange information among adjacent nodes across several layers.
The propagation process of GCN's layer $l$ can be expressed as:
\begin{equation} 
    \mathbf{H}^{(l+1)} = \sigma \left( \mathbf{D}^{-\frac{1}{2}} \left( \mathbf{A} + \mathbf{I}\right) \mathbf{D}^{-\frac{1}{2}} \mathbf{H}^{(l)} \mathbf{W}^{(l)} \right) \,,
\end{equation}
where
$\mathbf{H}^{(l)} \in \mathbb{R}^{N \times F^{(l)}}$ is the node feature matrix at layer $l$, with $N$ being the number of sessions
and $F^{(l)}$ being the feature dimension of each node.
$\mathbf{A} \in \mathbb{R}^{N \times N} $ is the adjacency matrix of the graph,
$\mathbf{I}$ is the identity matrix, 
and $\mathbf{D}$ is the degree matrix of $\mathbf{A}+\mathbf{I}$, i.e., \(\mathbf{D}_{ii} = \sum_j (A_{ij}\ + \delta_{ij})\).
$\mathbf{W}^{(l)} \in \mathbb{R}^{F^{(l)} \times F^{(l+1)}}$ is the weight matrix for layer $l$, which is learned during training.
We use ReLU activation function as $\sigma$.
For the final layer, we do not use an activation function, $\sigma$ is a simple linear transformation, but instead incorporate a pooling layer to control the output dimensionality.
So the final graph embedding is  $ \mathbf{G} = Pooling(\mathbf{H}^{(l+1)} ) $ .

\begin{figure}[t]
    \centering
    \includegraphics[width=2.6in]{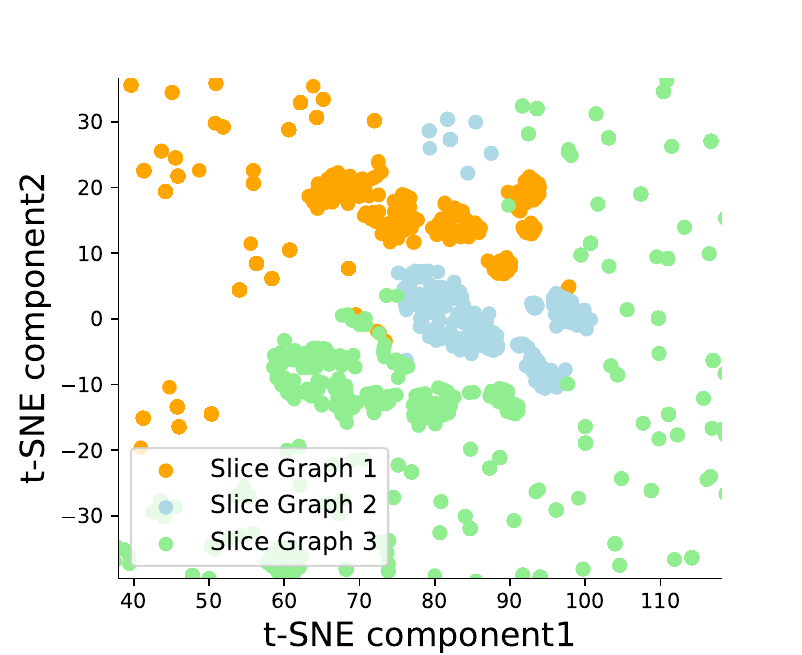}
    \caption{An example of slice-based graph embedding visualization.}\vspace{-0.2in}
    \label{fig:graph_embedding}
\end{figure}

To facilitate the interpretation of the structure within the high-dimensional graph data and better understand GCN's behaviors, we utilized a technique known as t-distributed stochastic neighbor embedding (t-SNE) \cite{t-sne} for visualizing the slice-based graph embeddings learned by the GCN for slicing and session typologies. 
Graph embeddings are projected into a two-dimensional space with the first and second components represented. 
For this example, we selected three slices and the dimensionality of our graph embeddings is 6. The visualization captures three distinct groups (Slice Graph 1, Slice Graph 2, and Slice Graph 3), each represented by a different color.
Each slice contains multiple sessions.

Fig.~\ref{fig:graph_embedding} shows sample outputs from the GCN. 
The resulting embeddings show that the GCN successfully captures the different features of the three slices, and the embedding effectively represents the features of each slice graph.
Scattered points outside the main clusters represent noise, outliers, or transitional data, such as sudden changes in channel or demand during a particular session, which are not well aligned with any particular slice. 
However, overall, the three slices can be clustered well.

\subsection{A DRL Framework}
\label{DRL training}

\paragraph{Key elements}
DRL is a type of online learning algorithm where an agent learns to make decisions by taking actions in an environment to maximize cumulative reward. 
It has the following four key elements:
\begin{itemize}[leftmargin=0.1in,itemsep=0in,topsep=0.0in]
\item 
\textbf{State.} 
The state representation includes the network state, channel conditions, session numbers, and UE activities. 
The original KPM and MAC data (see Table~\ref{tab:kpmmacdata}) are fed into the GCN module, and the output of GCN is the state of DRL, capturing a global view of network condition and traffic demand.

\item 
\textbf{Environment (RAN).} 
The RAN infrastructure serves as the environment for the DRL model. It receives resource slicing decisions from the DRL model, performs realistic resource allocation and measurements, and subsequently returns the reward metrics—including throughput, latency, and BLER for each session—along with the next state of the RAN.


\item 
\noindent\textbf{Action.} 
The action involves deciding the number of consecutive RBs allocated to each individual slice. In our approach, we define the output of the DRL model as a continuous action space. Since we only have access to the global CQI for each session, but lack specific CQI information for each slice associated with individual RBs and sessions, we represent the output as the ratio of RBs allocated to a slice relative to the total available BWP, constrained within the range [0, 1].

\item 
\noindent\textbf{Reward.} 
We use the inverse of the regret that we defined in Equation~\eqref{eq:opt1} as the reward. 
Specifically, we aim to maximize $u(t) - r(t)$, where $u(t)$ is the RB utilization efficiency and $r(t)$ is the regret over time. 



\end{itemize}
 
\noindent\paragraph{Training algorithm. } The DRL agent is trained by an Actor-Critic algorithm: Proximal Policy Optimization (PPO).
The actor and critic networks utilize a four-layer ($ h \times h $) Multi-Layer Perceptron (MLP) model where $h$ is the hidden size. 
We employ a Gaussian distribution to model the policy network's output. Specifically, the policy network outputs the mean ($\mu$) and standard deviation ($\sigma$) of the action distribution, where $\mu$ represents the most likely action to be taken and $\sigma$ reflects the confidence level of the policy network in that action. 
During training, the PPO algorithm optimizes $\mu$ and $\sigma$, enabling the policy network to sample the optimal resource allocation under the given state.

For the critic network in Fig.~\ref{fig:main_frame}, we use Generalized Advantage Estimation (GAE) \cite{schulman2015high} to calculate the advantage function, which determines how much better an action is compared to the action taken in state $s_t$.
Denote $\hat{A}_t$ as the advantage in iteration $t$. 
It can be written as:
\begin{equation}
    \hat{A}_t = (\gamma \lambda) \hat{A}_{t-1} + r + \gamma  V(s') - V(s) ,
    \label{RL_ADVANTAGE}
\end{equation}
where $V(s)$ is the value function to estimate the expected return (reward) from state $s$. 
$r$ denotes the reward.
The term $r+ \gamma V(s')$ is the Temporal Difference (TD) target, which represents the difference between the predicted value and the actual reward plus the discounted value of the next state. 
The discount factor $\gamma$ determines the importance of the future reward in the current decision-making process.
$\lambda$ is a weighting factor to smooth the estimation of the value function, thereby controlling the bias and variance in the estimates.

Based on the definition of $\hat{A}_t$, we train the policy network using the loss as follows:
\begin{align}
\mathcal{L}(\theta) 
= 
&-\mathbb{E}_t \left[ \min \left( R(\theta) \cdot \hat{A}_t,\;\;\; \text{clip}(R(\theta), \epsilon) \cdot \hat{A}_t \right) \right] \nonumber \\
&+ loss (V(s), d)\,,
\label{eq:RL_LOSS}
\end{align}
%
where $R(\theta)$ is the ratio of the new policy to the old policy. $\epsilon$ is the clipping parameter that limits the magnitude of policy updates. $loss (V(s), d)$ is the difference between $V$ and $d$.




Specifically, the training consists of the following two steps. 
\begin{itemize}[leftmargin=0.1in,itemsep=0in,topsep=0.0in]
\item 
\rev{\textbf{Step 1: Warm-up.} }
\rev{Since our approach is designed for Near-RT RIC, which requires scheduling decisions within a time window of 10 milliseconds to 1 second, pre-training plays a crucial role in stabilizing the DRL model. }
The primary goal of pre-training is to initialize the DRL agent with a reasonable policy that helps keep critical statistics---such as the mean and variance of actions---within the desired range, specifically between 0 and 1. 
In our slice resource control mechanism, the agent outputs the percentage of PRBs to be allocated to each network slice.

\item 
\textbf{Step 2: Online-learning.}
The agent engages in a cycle of trial and error, where taking actions and observing the resulting performance in the environment. The RAN environment provides feedback in the form of regret based on the measured performance metrics (throughput, latency, and BLER of each slice). As the agent continuously interacts with the environment, it updates its policy in real-time, learning to allocate resources more efficiently to meet the time-varying PRB demands of each slice.

\end{itemize}

\subsection{ Penalty Definition and Policy Updates}


\begin{figure}[t]
    \centering
    \includegraphics[width=2.5in]{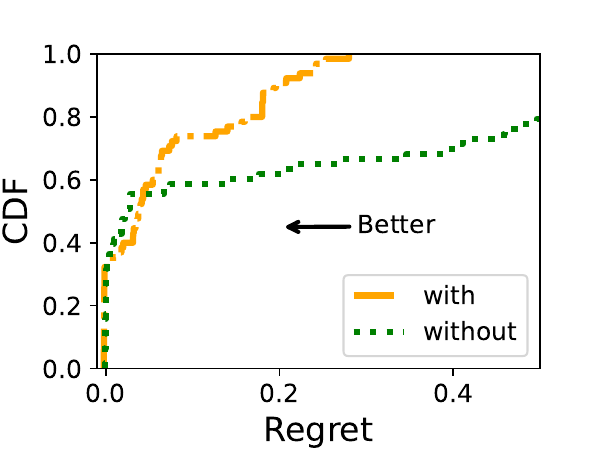} 
    \caption{Performance comparison of \pname with and without the proposed penalty mechanism.} 
    \label{fig:fine-runing}
\end{figure}

In our experiments, we observed dramatic regret fluctuation during short periods of time if there was a change in the session or channel states. 
This unstable regret fluctuation results in erroneous decisions on resource slicing, significantly degrading the network performance and disrupting the connection of some sessions. 
To illustrate this issue, let us consider a case where a session requires 5 RBs to maintain its stable connection. If the policy allocates fewer than 5 RBs, the session will be automatically disconnected, causing a large delay and a large regret. Additionally, if a session has a high demand for RBs but the policy cannot meet its demand, the session will also experience a significant delay during short periods of time. 

To mitigate this issue, \pname applies a large penalty for such cases, i.e., incorporating the performance degradation into the DRL memory but without updating the policy.
\rev{For the above case, we give a penalty $P$ and set the regret of this case to $P$, i.e., $r = P$.
Following the practice in \cite{yamagata2024safe}, 
we first normalized the per-step reward to be within a bounded range, and then empirically set $P$ to be a fixed fraction of the maximum achievable reward ($P = -0.2 R_{\max}$ in our experiments).}
To prevent the sharp decline in QoS performance, we apply a substantial penalty in the reward function during training, ensuring that poorly performing policies are not used in practice. 
We note that they are stored in the DRL memory solely as part of the training set.

To evaluate the effectiveness of this penalty mechanism, we conducted experiments on our O-RAN testbed (see \S\ref{sec:implement}) with 10 traffic sessions. 
\rev{We measure the regret value of \pname in two cases: with and without this penalty mechanism. 
Fig.~\ref{fig:fine-runing} presents our experimental results. }
It can be seen that, without the penalty mechanism, the regret is distributed over a large range.
In comparison, the penalty mechanism can significantly reduce regret, thus improving the quality of service of all sessions.

\section{Implementation}\label{sec:implement}
In this section, we present our implementation of \pname for evaluation. 
The source code of \pname is available on GitHub \cite{O-RANSlicingPeihao}.

 
\subsection{O-RAN Testbed}

\begin{figure} [t]
    \centering
    \includegraphics[width=\linewidth]{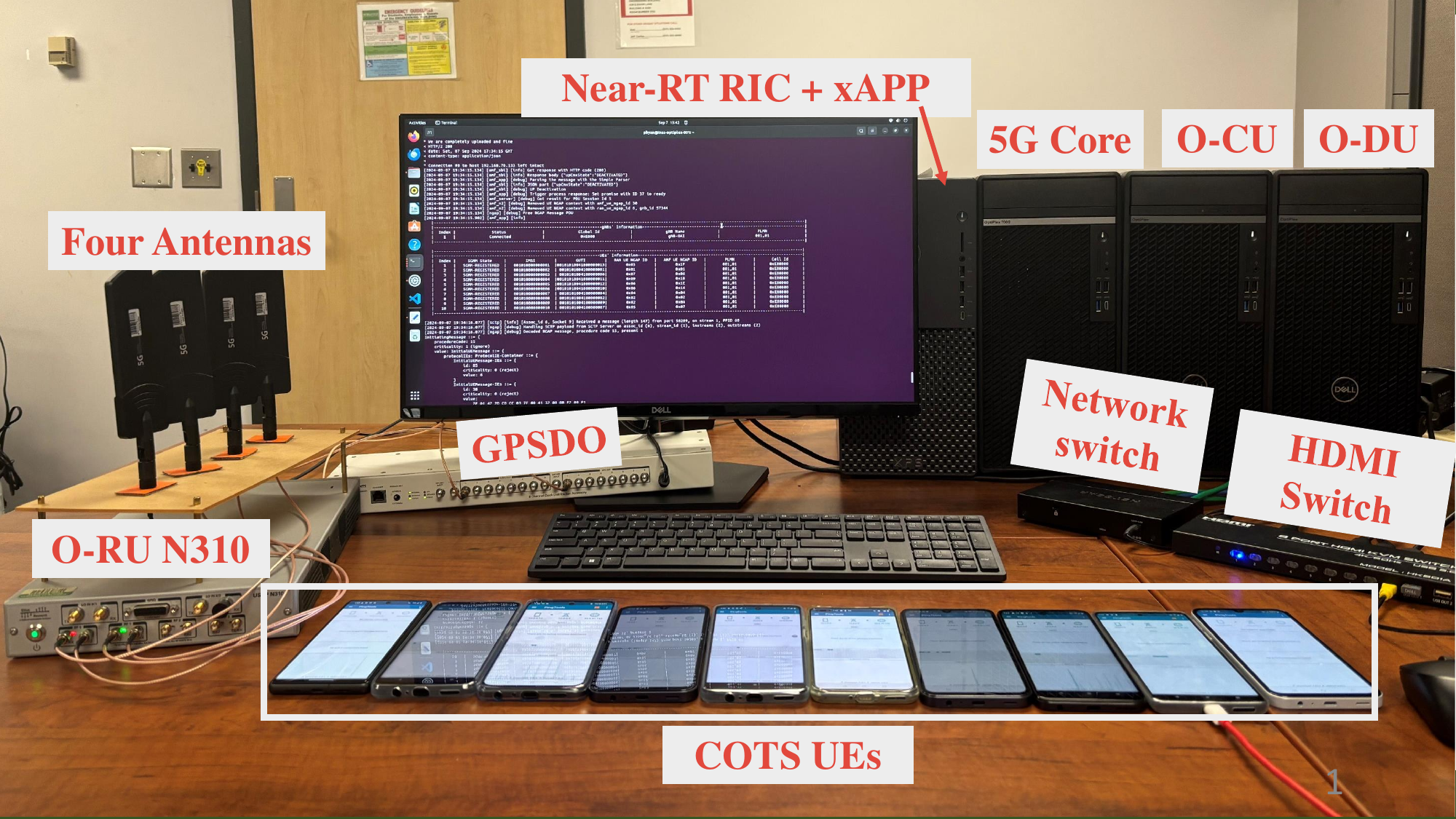}
    \caption{A picture of our O-RAN testbed with Near-RT RIC.} 
    \label{fig:testbed}
\end{figure}

\begin{table}
\renewcommand{\arraystretch}{1.15}
\footnotesize
\centering
\caption{Testbed hardware configuration.}
\begin{tabular}{ll}
\hline
\textbf{Servers (4×)}               & Intel(R) Core(TM) i9-14900K Core 24 (4x)\\ 
\textbf{External clock}               & GPSDO CDA-2990               \\ 
\textbf{Ethernet switch}            & Netgear GS308v3 \\ \hline
\textbf{Radio unit}                 & USRP N310 (4x4 MIMO)    \\
\textbf{Center frequency}                 &  TDD n78, 3319.68 MHz \\
\textbf{SubcarrierSpacing}                 &  30kHz \\
\textbf{UE brand}            & OnePlus, Google Pixel,        Motorola,  Samsung Galaxy      \\ \hline
\end{tabular} 
\label{tab:testbed_hardware}
\end{table}

\noindent\textbf{Testbed Hardware.} 
Fig.~\ref{fig:testbed} shows our O-RAN testbed, which includes 5G Core, O-CU, O-DU, O-RU and 10 COTS smartphones.
We implement the O-RAN system using 4 servers for O-DU, O-CU, Near-RT RIC, and 5G Core. Specific parameter settings are detailed in Table~\ref{tab:testbed_hardware}. The hardware configuration of this testbed is designed to support Internet access for COTS smartphones. Local network connection is provided by a Netgear GS308v3 Ethernet switch, which interconnects the 5G core, O-CU, O-DU, and O-RU components. The O-RU is set up on the USRP N310 device, supporting 4x4 MIMO configurations. 
USRP N310 uses external clock GPSDO CDA-2990 to improve its clock accuracy.
The radio units are tuned to a center frequency of 3319.68 MHz within the TDD n78 band, with a subcarrier spacing of 30 kHz. The testbed also includes 10 smartphones from OnePlus, Google Pixel, Motorola, and Samsung Galaxy, enabling a diverse range of testing scenarios and performance evaluation.
The n78 band was used for this project under an FCC Experimental License with Call Sign \#WA3XEP.

\noindent\textbf{5G Core Support.}
We have extended 5G core OAICN \cite{oai_cn} for this project. The AMF manages user access, authentication, and mobility, while the UPF handles user data traffic routing and quality of service. We have added various Slice Service Types (SST), which identify the slice type (e.g., eMBB, URLLC), and Slice Differentiators (SD) to the UPF database. When the NSSAI of a slice matches the corresponding SST and SD code in \verb|oai-cn5g/database/oai_db.sql|, the session is assigned to that slice.
 
\noindent\textbf{OpenAirInterface (OAI) Modifications.}
We use OAI \cite{openairinterface5g,villa2024x5g} 5G RAN for our experiments.
However, OAI does not support slicing. Therefore, we rewrote the downlink scheduler functions in \verb|oai/openair2/LAYER2/| \verb|NR_MAC_gNB| and slicing control E2 interface in \verb|oai/openair2/| \verb|E2AP/RAN_FUNCTION|. We define two attributes SD and SST for different slices, which serve as unique identifiers for the slices. When a UE connects to the 5G core network, the core database assigns each UE with an SST and SD, as well as 5G QoS Indicator (5QI) and priority level. The DU then assigns a Radio Network Temporary Identifier (RNTI) to each new UE to distinguish it.
Additionally, in the Single Slice model, all sessions utilize a single BWP, where the BWP size corresponds to the total number of PRBs. For \pname, we need to define a BWP for each slice, including its starting position and bandwidth size. The size can be obtained through \pname, while the starting position is determined by traversing through all of the PRBs.


\noindent\textbf{Near-RT RIC.}
\rev{We use Mosaic5g Flexric\cite{flexric, flexric_xapp} as our Near-RT RIC. Flexric contains E2 Node agent, Near-RT RIC and xApp. 
It provide a flatbuffers encoding/decoding scheme as alternative to ASN.1.}
We use SWIG as an interface generator to enable C/C++ and Python for the xApps.
We build our own xApp with E2AP v2.03 and KPM v2.03.

\subsection{\pname Implementation}
We implement \pname as an xApp within the Near-RT RIC. The Actor-Critic algorithm of the DRL agent is integrated into the PPO framework, supporting online training. The DRL environment is developed in Python for compatibility. \pname includes both the DRL model and the E2 interfaces for data collection and processing.
We have divided the task into two xApps: \textit{Monitor} xApp and \textit{AI-model} xApp. The Monitor xApp is responsible for interacting with the RAN to retrieve information (KPM and MAC data) and for sending updated slicing policies to the RAN. The AI-model xApp primarily focuses on model training and online learning.
By employing this multiprocess paradigm, \pname can promptly apply updated policies to the RAN while ensuring that the states acquired accurately reflect the implementation of the new policy actions.

The key parameter values of \pname are given in Table~\ref{tab:perameters} . 
The parameters include $\alpha$ (learning rate), $\gamma$ (discount factor), and $\lambda$ (weighting factor), which are set to 0.005, 0.95, and 0.2, respectively. Additionally, $e$ is the number of epochs and $b$ denotes the Mini-batch size. The table also presents values for three specific $\lambda$ parameters to calculate regret, which are used to adjust different aspects of the \pname.
\rev{The weights $\lambda_k^{[p]}$, $\lambda_k^{[d]}$, and $\lambda_k^{[r]}$ are selected through a small-scale grid search on the testbed, targeting a balance between throughput, delay, and reliability. The tuning objective is to ensure that no single QoS metric dominates the optimization, thereby avoiding bias toward a particular slice type. Specifically, we first normalize all QoS metrics to the range [0,1] based on their service-level targets, then adjust the weights such that each term contributes similarly to the overall regret in typical traffic conditions.}

\begin{table}[t]
\renewcommand{\arraystretch}{1.15}
\scriptsize
\centering
\caption{Parameter value of \pname.}
\begin{tabular}{|c|c|c|c|c|c|c|c|c|}
\hline
$\alpha$ & $\gamma$ & $\lambda$ & e & b & $\lambda_{k}^{\mathrm{[p]}}$ & $\lambda_{k}^{\mathrm{[d]}}$ & $\lambda_{k}^{\mathrm{[r]}}$ & C \\\hline
0.005 & 0.95 & 0.2  & 16 & 10 & 1 & 0.8 & 2 & 0 \\ \hline
\end{tabular} 
\label{tab:perameters}
\end{table}


\begin{figure}[t]
    \centering
    \includegraphics[width=\linewidth, trim=40 20 60 0, clip]{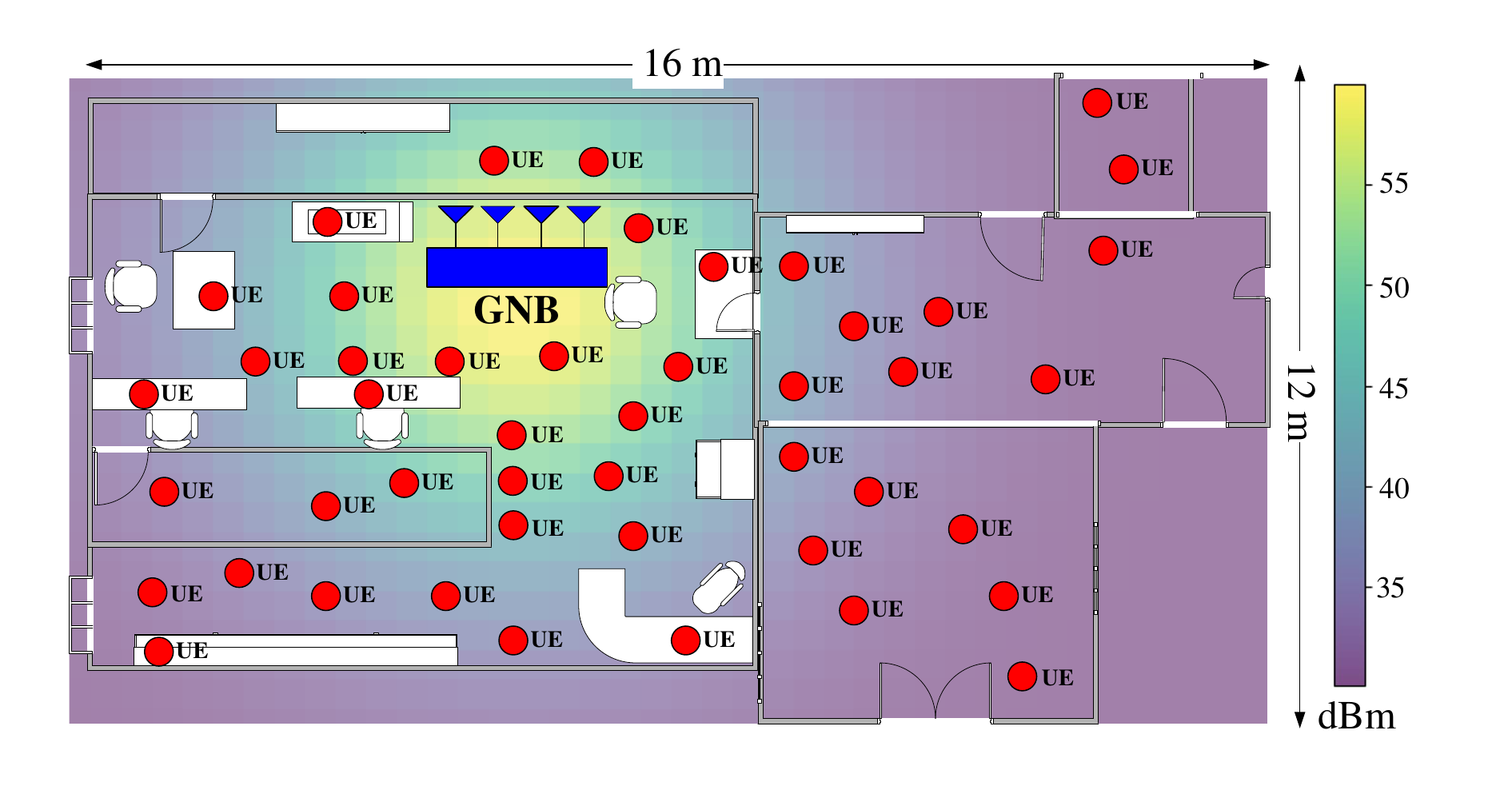}\vspace{-0.1in}
    \caption{gNB and UE locations as well as PHR values.}\vspace{-0.15in}
    \label{fig:signal_coverage}
\end{figure}

\subsection{Experimental Setup}
Fig.~\ref{fig:signal_coverage} shows our experimental scenarios and the 5G signal coverage.
The red circles indicate the possible location of UEs.
The color gradient ranges from yellow (indicating higher signal strength) to purple (indicating lower signal strength), representing the signal intensity (in dBm) across the entire area. 
The signal strength varies with distance from the gNB, displaying stronger signals near the gNB and weaker signals in more distant rooms.


\section{Experimental Evaluation}\label{sec:experiment}
In this section, we conduct extensive OTA experiments to evaluate the performance of \pname in realistic scenarios.

\subsection{Performance Baselines}
\label{baselines}

In addition to \pname, we have implemented the following three resource slicing policies in our O-RAN testbed and use them as the baselines for performance comparison. 

\begin{itemize}[leftmargin=0.1in,itemsep=0in,topsep=0.0in]
\item 
\textbf{Single Slice Policy:} 
We assume an ideal situation in which session traffic demand and channel states are known, allowing us to compute the optimal bandwidth allocation scheme. In this scenario, all sessions are scheduled together in a single slice using a proportional fair scheduler, which is widely employed by today's incumbent base stations.

\item 
\textbf{NVS Policy \cite{NVS}:} 
This is a dynamic network slicing algorithm for WiMAX, providing slice-level QoS guarantees by multiplexing slices over time. 
Each slice requests an aggregate throughput (for all users). 
The NVS controller tracks each slice's throughput. 
In each time interval (e.g., 10 ms), it computes the priority of each slice, which is defined as the ratio of requested throughput to average throughput.
Then, it selects the slice with the highest priority. 
Our implementation is based on the source code in \cite{FlexSliceGlobecom2023}.

\item 
\textbf{Zipper Policy \cite{app_based}:} It casts the problem as a model predictive controller, and explicitly tracks the network dynamics of each user to simplify the search space. A primitive that estimates if there is bandwidth available to accommodate an incoming app helps improve resource efficiency. It assumes all the app's  Service Level Agreement (SLA) (e.g., traffic demand) and channel states are known.



\item 
\textbf{ \rev{IQRA Policy \cite{mhatre2024intelligent}:}} 
\rev{It targets QoS-aware intra-slice resource allocation in O-RAN and proposes a DQN-based framework to optimize resource distribution for eMBB and URLLC slices. The framework introduces user association parameterization. By dynamically adjusting wireless resource allocation according to traffic demand and QoS constraints.}

\begin{figure}[t]
    \centering
    \includegraphics[width=\linewidth, trim=20 0 80 0, clip]{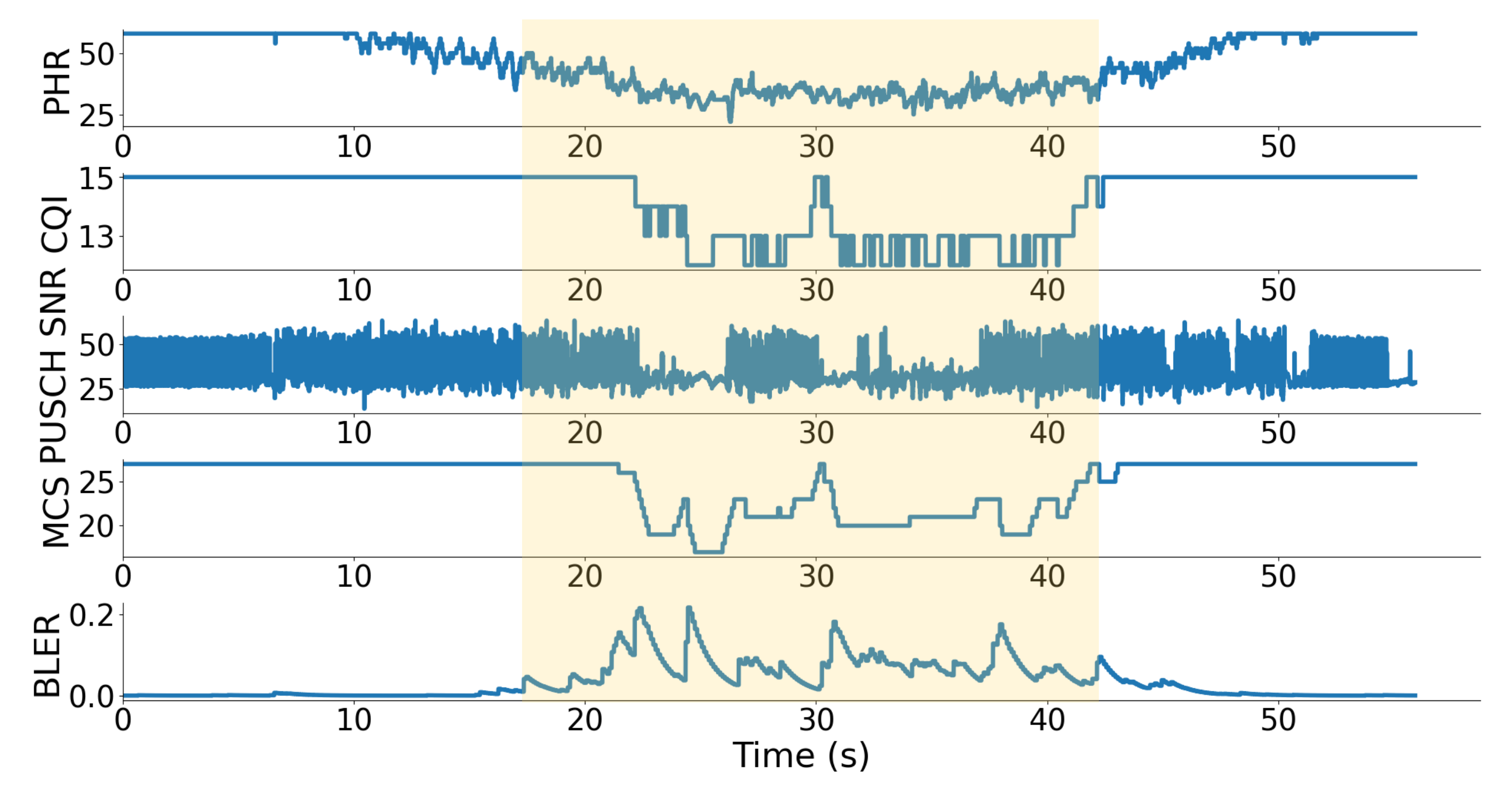}\vspace{-0.1in}
    \caption{The changes of PHR, CQI, PUSCH SNR, MCS, BLER of a phone when it moves.}\vspace{-0.15in}  
    \label{fig:uemovecqi}
\end{figure}

\item 
\textbf{\rev{ GNN+TD3 Policy \cite{liu2024achieving}:}} 
\rev{It proposes a network slicing-based framework to enhance energy efficiency in 5G access networks. A GNN predicts slice workloads from dynamic traffic and guides the frequency-domain resource allocation, while a TD3 strategy with CPU C-states optimizes load balancing.}

\end{itemize}

\subsection{Case Studies}
\label{subsec:case_study}

Before delving into the extensive experimental evaluation, we first take a close look at some components of the O-RAN system and the proposed \pname through case studies.

\begin{figure}[t]
    \centering
    \includegraphics[width=3.6in, trim= 0 0 0 0, clip]{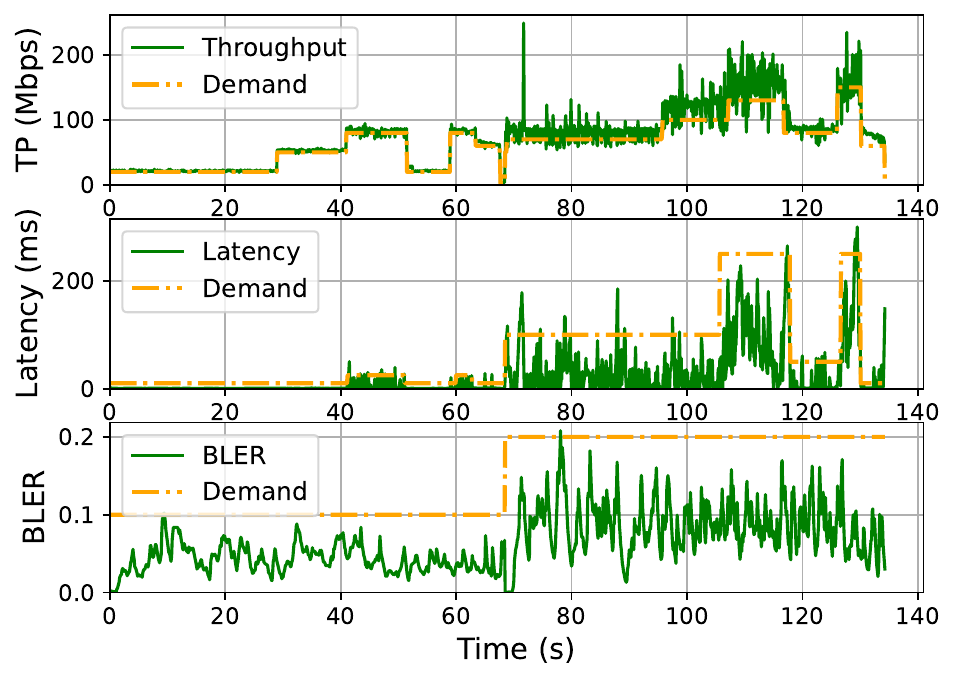}\vspace{-0.1in}
    \caption{\rev{Achieved throughput (TP), latency and BLER of a dynamic traffic session \textit{versus} its demands.}}\vspace{-0.15in}
    \label{fig:demand_track}
\end{figure}

\noindent\textbf{Channel Dynamics.}
We first measure experimental data to verify the session dynamics over time by collecting the PHR, CQI, PUSCH SNR, MCS, and BLER of a downlink traffic session from gNB to a smartphone. 
The smartphone is carried by a person who is working in her office in a routine fashion.
Fig.~\ref{fig:uemovecqi} presents our experimental results. 
It can be seen that the measured metrics indeed change over time, indicating the dynamics of DRL state (an aggregate of the above metrics and others) over time and the necessity of adaptation on resource slicing.


\noindent\textbf{Demand Dynamics.}
\rev{For a UE's application, a traffic session has a latency requirement between 20ms and 250ms, a BLER demand of 0.1 to 0.2, and a time-varying throughput demand ranging from 20 Mbps to 180 Mbps.} We implement this traffic session from the gNB to a smartphone using the \verb|iperf| command. We intentionally change the throughput demand over time to observe the reaction of \pname.

\rev{
Fig.~\ref{fig:demand_track}(top) presents the achieved throughput alongside its dynamic demand. The system consistently meets the increasing demand, which rises in distinct steps. Notably, even during sharp demand increases, such as at approximately 70 seconds (from 60 Mbps to 100 Mbps) and 100 seconds (from 100 Mbps to over 150 Mbps), the achieved throughput stabilizes and closely aligns with the new demand after a short time period of variability. This demonstrates \pname's ability to adapt its resource allocation strategies in real-time in the face of dynamic throughput demands. The minor fluctuations during these transition periods reflect the model's online learning process as it refines its decision-making to meet the new requirements.}

\rev{
Fig.~\ref{fig:demand_track}(middle) shows the achieved latency and its time-varying demand. The latency demand, initially constant, increases significantly at the 70-second mark, coinciding with the first major throughput demand spike. However, even with the latency increase, the system effectively manages to keep latency below the specified 200ms demand. The system dynamically trades off latency to achieve a higher throughput, but still within the demand range.}

\begin{figure}
    \centering
    \includegraphics[width=3.3in, trim=0 0 0 0, clip]{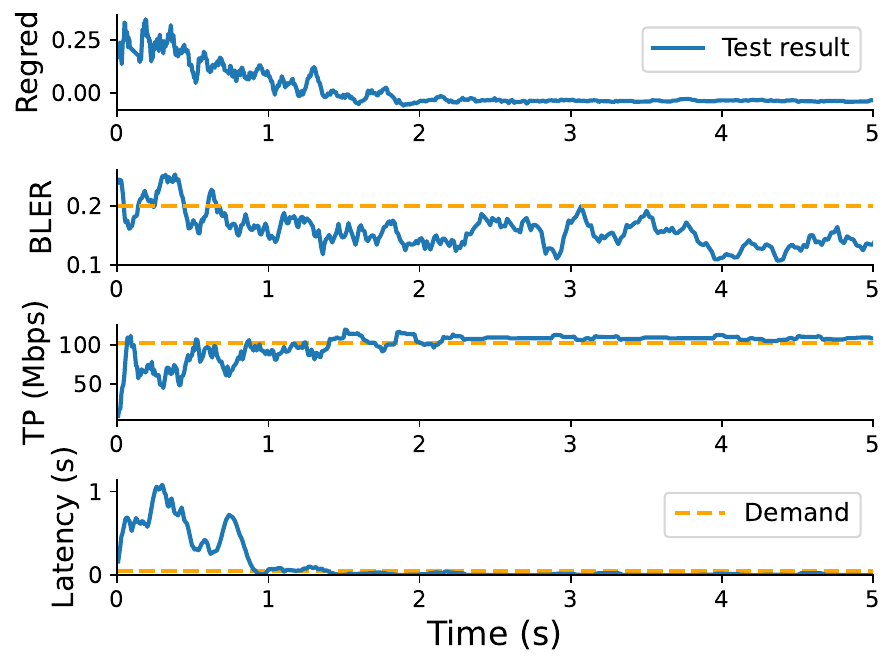}\vspace{-0.1in}
    \caption{\rev{Adaptability studies of \pname when the network has a new traffic session.}}\vspace{-0.1in}
    \label{fig:all_converge}
\end{figure}

\rev{
Fig.~\ref{fig:demand_track}(bottom) shows the achieved BLER and its time-varying demand. The BLER demand is set to 10\% at the beginning and then set to 20\% at the 70 second. 
This demonstrates that \pname can maintain the QoS requirement for BLER. While there are minor fluctuations, the system successfully keeps the BLER below the specified demand, which is crucial for ensuring reliable communication. The stable and low BLER performance, even during the periods of high throughput demand, indicates \pname's effectiveness in balancing various QoS metrics simultaneously.}
\noindent
\rev{
\textbf{DRL Adaptability.}
We now examine how \pname responds to a newly-arrived traffic session from a UE. In the 5G core network, we generate 100 Mbps of data traffic for a Samsung phone, setting its latency and BLER demands to 30 ms and 0.2, respectively. Fig.~\ref{fig:all_converge} displays our observations on regret, throughput, latency, and BLER from the moment the traffic session is initiated.
It is evident that the DRL model stabilizes its performance within approximately 2 seconds. Notably, the new traffic session initially experiences high latency (up to 1 second), but this elevated latency lasts for only about 1 second. Afterward, it decreases rapidly and meets the latency demand.
This showcases the adaptability of our DRL model in the face of data traffic dynamics.
}

\subsection{\rev{Slice Weight Analysis}}

\rev{For notational simplicity, we first define a base weight vector as $\vec{\lambda} = [r^{\mathrm{[p]}}, r^{\mathrm{[d]}}, r^{\mathrm{[r]}}] = [1,\; 0.8, \;2]$. }

\begin{figure}[t]
    \centering
    \begin{subfigure} [b]{0.75\linewidth}
        \centering
        \includegraphics[width=\linewidth]{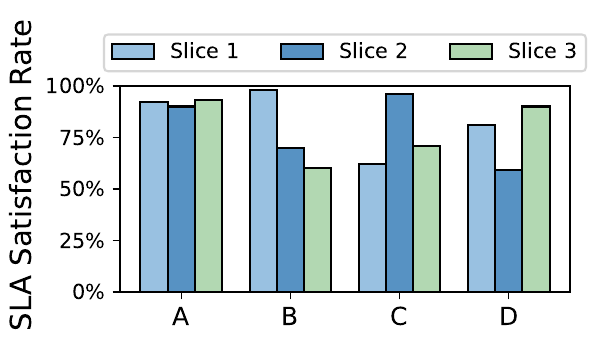}
    \end{subfigure}
    \caption{\rev{Performance comparison of different weights in medium traffic case (traffic demand: 20--180 Mbps).}}\vspace{-0.2in}
    \label{fig:for_each_slice}
\end{figure}

\noindent
\rev{\textbf{ Inter-Slice Weight Difference.} 
We evaluate four different cases of weight vector combinations across three separate slices. 
\textbf{Case A:}
All slices use $\vec{\lambda}$.
\textbf{Case B:}
Slice 1 uses $2\vec{\lambda}$ and other slices use $\vec{\lambda}$.
\textbf{Case C:}
Slice 2 uses $2\vec{\lambda}$ and other slices use $\vec{\lambda}$.
\textbf{Case D:}
Slice 3 uses $2\vec{\lambda}$ and other slices use $\vec{\lambda}$.
For each case, we conduct experiments and recorded the SLA satisfaction rate for every slice, with the results detailed in Fig.~\ref{fig:for_each_slice}. 
The results clearly indicate that the weights in the regret optimization function directly affect each slice’s performance. When all slices are assigned equal weights (Case A), their SLA satisfaction rates remain uniformly high and nearly identical. In contrast, when the weight of a single slice is doubled (Cases B, C, and D), that slice consistently achieves an almost perfect satisfaction rate, while the satisfaction rates of the remaining slices decline significantly. This highlights a fundamental trade-off: prioritizing the performance of one slice inevitably comes at the expense of others.
}

\noindent
\rev{\textbf{Intra-Slice Weight Perturbation. }To assess the robustness of our parameters, we next conducted a sensitivity analysis. Starting with our base parameter vector, $\vec{\lambda} = [1, 0.8, 2]$, we introduced a small perturbation of $\delta = 0.2$. We then tested the robustness by individually varying each component—$\lambda_{k}^{\mathrm{[p]}}$, $\lambda_{k}^{\mathrm{[d]}}$, and $\lambda_{k}^{\mathrm{[r]}}$—by adding and subtracting $\delta$.}

\begin{figure*}
\begin{minipage}{\linewidth}
    \centering
    \begin{subfigure} [b]{\linewidth}
        \centering
        \includegraphics[width=\linewidth]{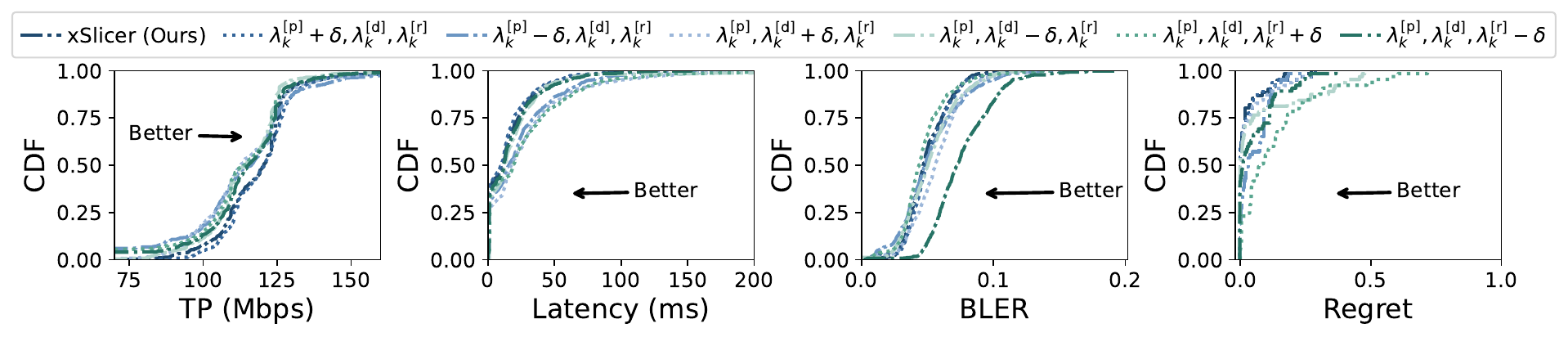}
    \end{subfigure}
    \caption{\rev{Performance comparison of different weights in medium traffic case (traffic demand: 60-180 Mbps).}}
    \label{fig:lambda_compare}
\end{minipage} 
\end{figure*}

\rev{This analysis demonstrates that our choice of weights is robust to small perturbations.
As shown in Fig.~\ref{fig:lambda_compare}, even when we slightly adjust a single parameter$\lambda^p_k$, $\lambda^d_k$, or $\lambda^r_k$ by adding or subtracting a small value of $\delta = 0.2$ from our base parameter vector, the overall performance remains largely unaffected. 
The curves for the perturbed parameters closely track the curve for our original parameters, indicating that minor deviations in our weight selection have little to no impact on the final results. This confirms that our chosen weights are not overly sensitive and provide a stable and reliable solution.}

\subsection{Hyperparameter Analysis} \label{subsec:Hyperparameter_Analysis}

Recall that \pname consists of two key components: GCN and DRL. In this section, we study their hyperparameters to identify the optimal parameter configuration for \pname. We connect 10 smartphones to the O-RAN, with each smartphone having one traffic session with dynamic throughput demands. We repeat the experiments with various hyperparameter settings.

\begin{figure}[b]
    \centering
    \vspace{-0.2in}
    \includegraphics[width=2.6in, trim=0 0 0 0, clip]{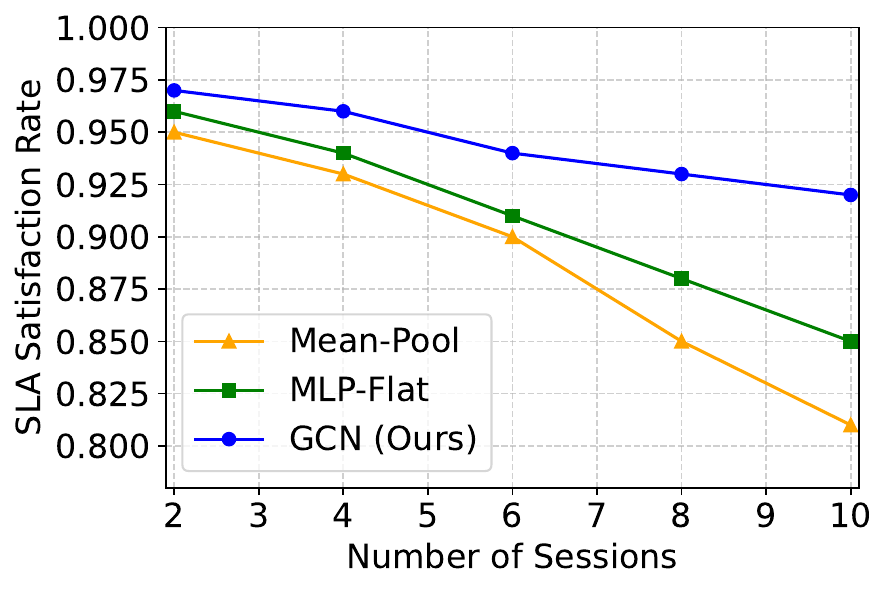}\vspace{-0.1in}
    \caption{\rev{Effectiveness of GCN.}}
    \label{fig:compare_gcn}
\end{figure}


\begin{figure*}
     \centering
     \begin{minipage}[b]{2.2in}
         \centering
        \includegraphics[width=2.2in]{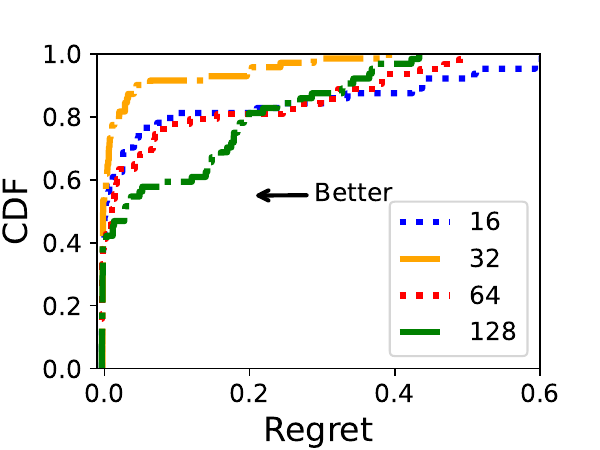}
        \caption{Impact of hidden size in MLP on DRL model.}
         \label{fig: impact of RL}
     \end{minipage}
     \hfill
     \begin{minipage}[b]{2.2in}
         \centering
         \includegraphics[width=2.2in]{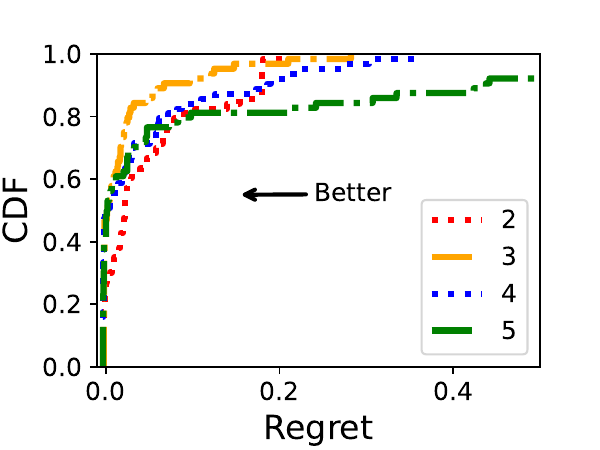}
         \caption{Impact of the number of GCN layers.}
         \label{fig:GCN layers}
     \end{minipage}
     \hfill     
     \begin{minipage}[b]{2.2in}
         \centering
         \includegraphics[width=2.2in]{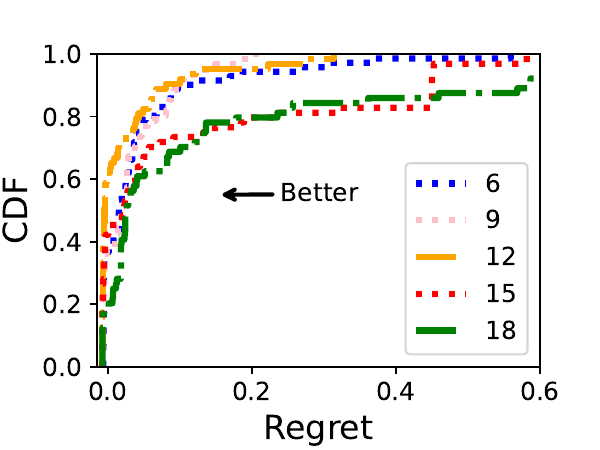}
         \caption{Impact of GCN's output embedding size.}
         \label{fig:Embedding sizes}
     \end{minipage}
    \label{fig:Impect of hyperparameters}
\end{figure*}

\noindent
\rev{\textbf{GCN Effectiveness.}
We compare the SLA satisfaction rate of our proposed GCN model against MLP-Flat and Mean-Pool as the number of user sessions increases. As shown in Fig.~\ref{fig:compare_gcn}, although all models experience performance degradation under higher load, our GCN consistently achieves the highest satisfaction rate, highlighting its superior ability to preserve service quality in congested network conditions.}
\rev{The SLA satisfaction rate is a key performance indicator (KPI) that measures the proportion of time a service meets its contractually defined targets, such as uptime, latency, or throughput. It reflects service reliability by capturing the percentage of successful service delivery instances relative to the total over a given period. A higher rate indicates stronger performance and greater compliance with the service agreement.}

\noindent
\textbf{DRL's Hidden Layer Size.}  
Fig.~\ref{fig: impact of RL} shows the changes in regret when DRL employs 16, 32, 64, and 128 hidden neurons in its MLP. The results indicate that (i) \pname exhibits high regret and significant fluctuations at the beginning; (ii) \pname converges to a similarly low regret across all configurations. Despite this ultimate convergence, the convergence speeds differ. Among the tested configurations, the DRL with 32 hidden layer neurons demonstrates the best convergence speed. Increasing the hidden layer size further may enhance convergence speed, but it will also increase computational load.


\noindent
\textbf{Number of GCN Layers.}  
We conducted similar experiments with the GCN varying its number of layers from 2 to 5. Fig.~\ref{fig:GCN layers} presents the distribution of regret in the O-RAN for different GCN layer configurations. It is evident that \pname achieves the best regret performance with the GCN consisting of 3 layers. In this slicing graph structure, which does not exhibit complex relationships, a shallow GCN is sufficient for capturing fundamental feature representations. However, increasing the number of layers may lead to slow convergence, resulting in increased response times in near-real-time slicing and, consequently, greater regret.


\noindent
\textbf{GCN's Embedding Sizes.}  
In this test, we fixed the DRL's hidden layer size at 32 and the GCN's layer count at 3. We examine the impact of embedding dimension size on demand satisfaction. Fig.~\ref{fig:Embedding sizes} presents our experimental results. 
We can observe that the embedding size 12 achieves the best performance. 
If the embedding size is excessively large, the GCN may struggle to effectively extract and integrate features, potentially confusing the DRL model. Conversely, an excessively small embedding size may lead to the loss of critical features in the output.

\begin{figure*} [t]
\begin{minipage}{\linewidth}
    \centering
    \begin{subfigure} [b]{\linewidth}
        \centering
        \includegraphics[width=\linewidth]{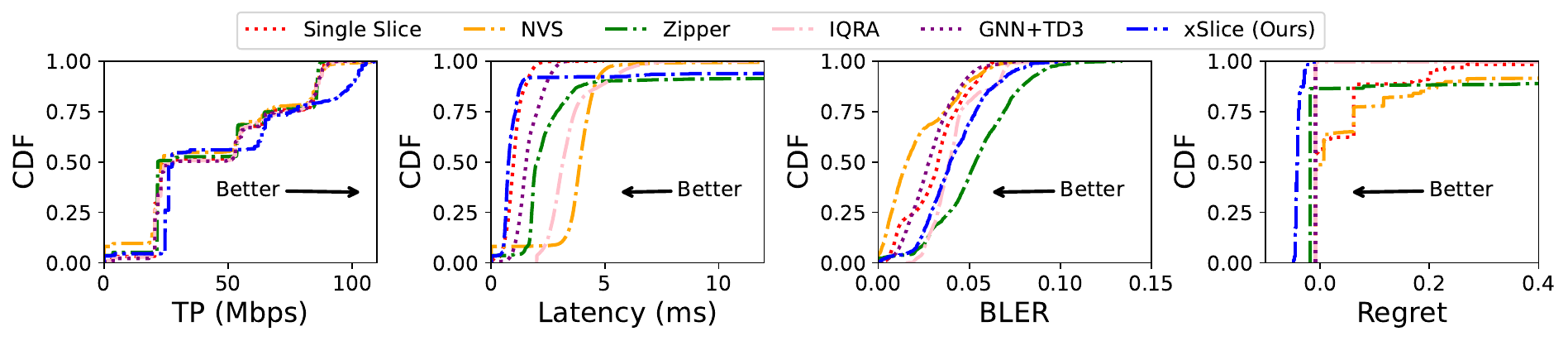}
        \label{fig:50_throuput}
    \end{subfigure}\vspace{-0.1in}
    \caption{\rev{Performance comparison of \pname and existing approaches in light traffic case (traffic demand: 20--80 Mbps).}}
    \label{fig:20M_80Mbps}
\end{minipage}
\begin{minipage}{\linewidth}
    \centering
    \begin{subfigure} [b]{\linewidth}
        \centering
        \includegraphics[width=\linewidth]{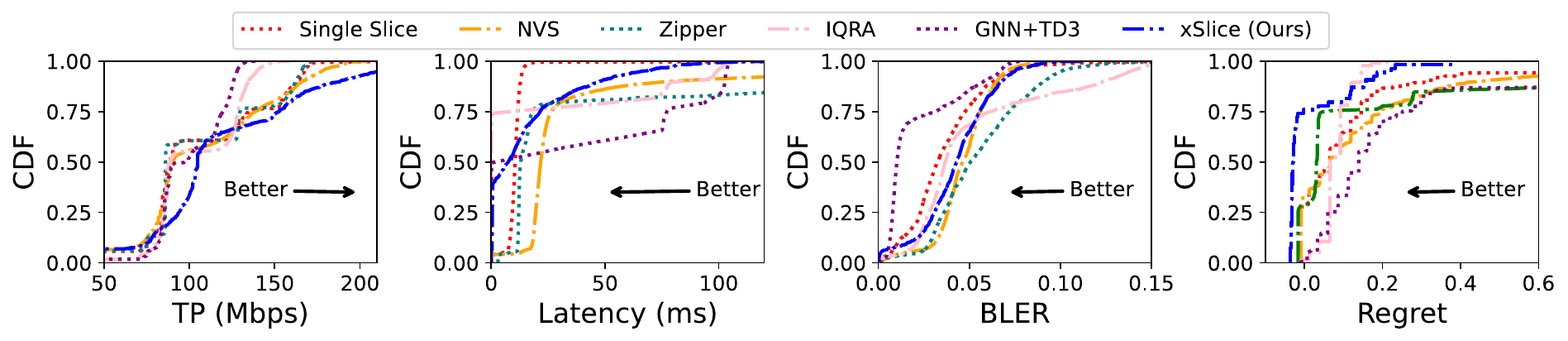}
        \label{fig:100_throuput}
    \end{subfigure}\vspace{-0.1in}
    \caption{\rev{Performance comparison of \pname and existing approaches in medium traffic case (traffic demand: 80--160 Mbps).}}
    \label{fig:80M_150Mbps}
\end{minipage}
\begin{minipage}{\linewidth}
    \centering
    
    \begin{subfigure} [b]{\linewidth}
        \centering
        \includegraphics[width=\linewidth]{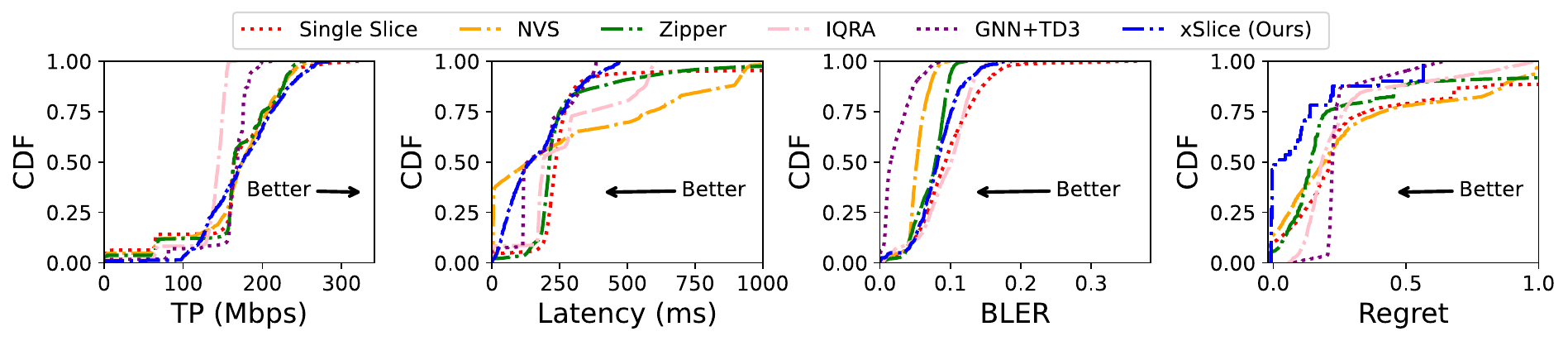}
        \label{fig:150_throuput}
    \end{subfigure}\vspace{-0.1in}
    \caption{\rev{Performance comparison of \pname and existing approaches in intensive traffic case (traffic demand: 160--210 Mbps).}}
    \label{fig:150M_210Mbps}
\end{minipage}   
\end{figure*}

\subsection{Extensive Evaluation in Different Traffic Cases}
\label{subsec:Comprehensive_Evaluation}

In this section, we conduct extensive experiments to evaluate the performance of \pname. In our experiments, the O-RAN serves 10 smartphones, each with one demand-dynamic traffic session. 
\rev{We use Single Slice, NVS, Zipper, IQRA and GNN+TD3 as performance baselines. }
In our experiments, we consider three cases: light traffic, medium traffic, and intensive traffic.

\noindent\textbf{Light Traffic Case.}
In this case, the 5G core network generates light traffic ranging from 20 Mbps to 80 Mbps to each of the 10 smartphones. 
\rev{Fig.~\ref{fig:20M_80Mbps} presents the comparison of \pname and existing resource slicing policies (Single Slice, NVS, Zipper, IQRA and GNN+TD3) in terms of their throughput, latency, reliability (BLER), and overall regret. }

In terms of \textit{throughput}, all policies perform similarly.
This is intuitive given the light traffic conditions. 
\rev{The RAN can easily meet the throughput demand even with a sub-optimal policy. 
Regarding \textit{latency}, \pname is the best policy, with Single Slice closely following. }
This is because the resource is more than sufficient to meet the demand under all slicing policies. 
For \textit{BLER}, NVS shows the best performance. 
For the \textit{regret} calculated based on the measured throughput, latency, and BLER, \pname outperforms all the other three policies, as shown in the figure. 
This is because \pname optimally balances them to achieve an overall regret performance. 
Recall that \textit{regret} is defined as the weighted throughput, latency, and BLER (reliability) deficits, allowing network operators to easily prioritize throughput, latency, or reliability by adjusting the corresponding weights.
\rev{Compared to the DRL-based baseline models IQRA and GNN+TD3, \pname achieved lower regret in low-traffic scenarios. However, IQRA and GNN+TD3 also significantly outperformed the other algorithms. This demonstrates that reinforcement learning-based methods are able to adapt to changing demand.}

\noindent\textbf{Medium Traffic Case.}
In this case, the 5G core network generates traffic ranging from 80 Mbps to 160 Mbps for each of the 10 smartphones.
Fig.~\ref{fig:80M_150Mbps} compares \pname with existing resource slicing policies in terms of throughput, latency, BLER, and regret.
\pname demonstrates a clear advantage over other policies in minimizing regret. Leveraging its GCN and DRL modules, \pname effectively learns the key features of the traffic sessions to achieve the best regret performance. 
Moreover, it slightly outperforms other policies in terms of throughput and latency. 
\rev{In particular, while IQRA and GNN+TD3 also employ reinforcement learning for slice optimization, they fall behind \pname in regret performance. This is because their designs either lack a fine-grained graph representation of session dynamics (IQRA) or are limited by suboptimal reward shaping (GNN+TD3), making it difficult to balance the trade-offs among KPIs under moderate resource pressure.}

\noindent\textbf{Intensive Traffic Case.}
In this scenario, the 5G core network generates intensive traffic, ranging from 160 Mbps to 220 Mbps, for each of the 10 smartphones.
Fig.~\ref{fig:150M_210Mbps} depicts the distribution of the measured throughput, latency, BLER, and regret under different resource slicing policies.
As expected, latency is significantly high in this scenario because the O-RAN lacks sufficient resources to meet the demands of the UEs.
Experimental results indicate that latency can reach as high as one second in certain cases.
Nevertheless, \pname consistently outperforms other policies tested, particularly in terms of latency and regret.
Notably, a significant reduction in regret is evident in the experimental results presented in the figure.
\rev{Under intensive traffic, both IQRA and GNN+TD3 show degraded performance, as their policies struggle to generalize when the O-RAN resources are severely constrained. In contrast, \pname leverages GCN-based feature extraction to better capture traffic correlations, enabling more efficient resource allocation and lower regret.}

\begin{table}[t]
\renewcommand{\arraystretch}{1.12}
\caption{Performance summary of \pname and other policies.}
\resizebox{\linewidth}{!}{
\begin{tabular}{|l|c|c|c|c|}
\hline
& TP (Mbps) $^\uparrow$ & Latency (ms) $^\downarrow$~~ & ~~BLER $^\uparrow$~~~& Regret$^\downarrow$~~~ \\\hline
Single Slice  &105.4 &245.9 &0.075 & 0.436\\ \hline
NVS  &106.6 &247.1 &0.041 & 0.493\\ \hline
Zipper & 105.5 & 116.3 & 0.061 & 0.244\\ \hline
IQRA  &106.8 &152.3 &0.052 & 0.298\\ \hline
GNN+TD3 & 105.5 & 137.2 & 0.091 & 0.329\\ \hline
\pname & 108.1 & 63.5& 0.063 & 0.079\\ \hline
\end{tabular} 
}
\label{tab:average_performance_value}
\end{table}

\noindent\textbf{Evaluation Summary.}
Table \ref{tab:average_performance_value} summarizes the performance of the four policies in terms of their average throughput, latency, BLER, and regret values. Numerically, \pname demonstrates superior performance in terms of regret, throughput, and latency compared to existing policies. Notably, when compared to Zipper (the state-of-the-art resource slicing policy), \pname reduces performance regret by 67\%.
 


\subsection{Inference Time}
\begin{figure}
    \centering
    \includegraphics[width=3.5in]{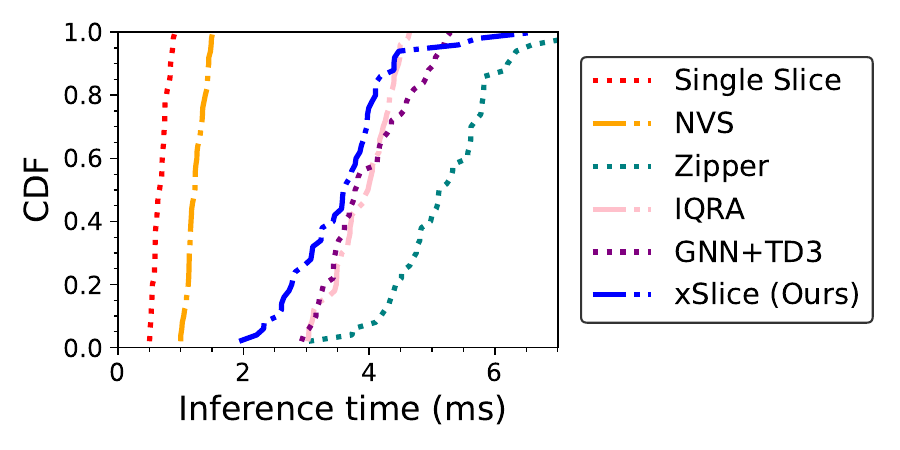}
    \caption{\rev{Inference time of \pname and other approaches.}}
    \label{fig:comp_time}
\end{figure}

\rev{ \pname is implemented as an xApp running within the Near-RT RIC, using the E2 interface to communicate with the RAN. Since the delay requirement for Near-RT operations ranges from 10 ms to 1 s, we evaluate the inference time of \pname under realistic operational conditions to ensure compliance. As shown in Fig.~\ref{fig:comp_time}, the resource slicing decision is completed within 4 ms in nearly all cases---well below the Near-RT latency bound. This ensures that decisions can be delivered to the RAN within 10 ms, making \pname suitable for practical RAN control. }

\rev{
Fig.~\ref{fig:comp_time} further compares inference times across baselines. }
\rev{\pname exhibits a longer computation time compared to Single Slice and NVS, since the latter are model-based slicers with deterministic resource allocation formulas, while \pname is learning-based and depends on a neural network pipeline for decision-making. 
Nevertheless, \pname achieves slightly shorter computation times than Zipper, IQRA, and GNN+TD3. This advantage stems from two key factors: first, \pname operates at the slice level rather than the per-user level, which significantly reduces the decision space and the input/output dimensions; second, it employs GCN-based feature extraction, where the GCN module aggregates per-UE information into compact per-slice embeddings in a single forward pass, thereby avoiding repeated per-user processing and enabling the policy network to handle a fixed-size input irrespective of the number of users.
}


\section{Conclusion}
In this paper, we introduced \pname, an online-learning 5G slicing xApp for O-RANs. \pname consists of two key components: a GCN module and a DRL framework. It defines a regret-based objective function that considers weighted throughput, latency, and reliability for all sessions in an O-RAN system, enabling online optimization for resource allocation.
To address the challenges associated with network dynamics, \pname employs the multi-layer GCN to extract feature embeddings from all sessions within a resource slice. This approach effectively integrates dynamic information, enhancing decision-making for resource allocation.
To facilitate online training and achieve Near-RT control of resource allocation, \pname adopts PPO, a DRL algorithm, to minimize performance regret. We have implemented \pname on a 5G O-RAN testbed and evaluated its performance in real-world O-RAN systems.
OTA Experimental results demonstrate that \pname can reduce the performance regret by 67\% compared to the state-of-the-art solutions.





\bibliographystyle{ieeetr}

%

\vspace{-0.2in}
\begin{IEEEbiography}
[{\includegraphics[width=1in,height=1.25in,clip,keepaspectratio]{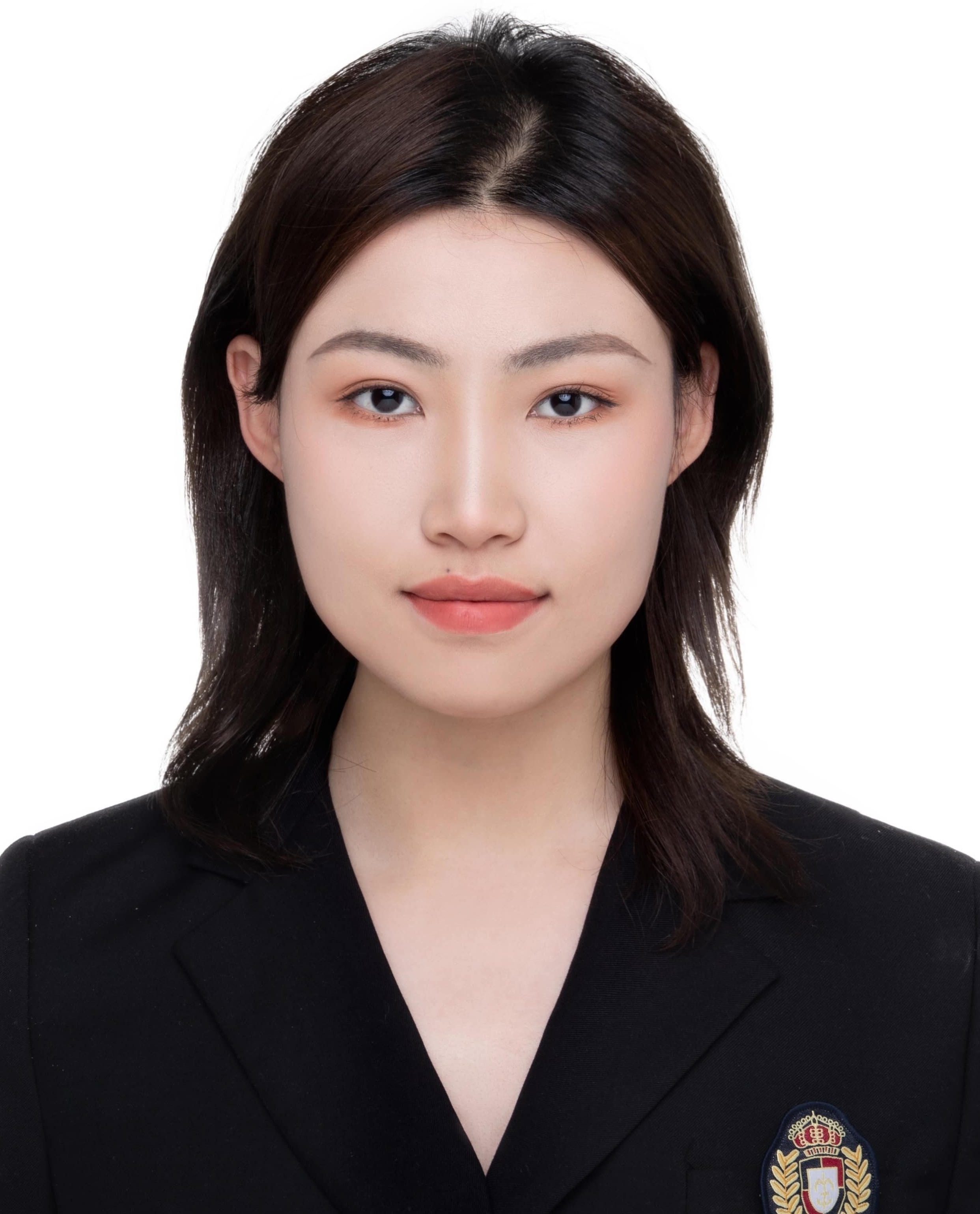}}]
{Peihao Yan} 
is currently a PhD student in the Department of Computer Science and Engineering at Michigan State University (MSU), East Lansing, MI. She received her B.E. degree in communication engineering in 2020 and her M.S. degree in Computer Science and Technology in 2023 from China University of Petroleum (East China), Qingdao, China. Her current research interests include wireless networking and communications, with emphasis on 5G systems, O-RAN, and AI-driven learning algorithms.
\end{IEEEbiography}
 
\vspace{-0.2in}
\begin{IEEEbiography}
[{\includegraphics[width=1in,height=1.25in,clip,keepaspectratio]{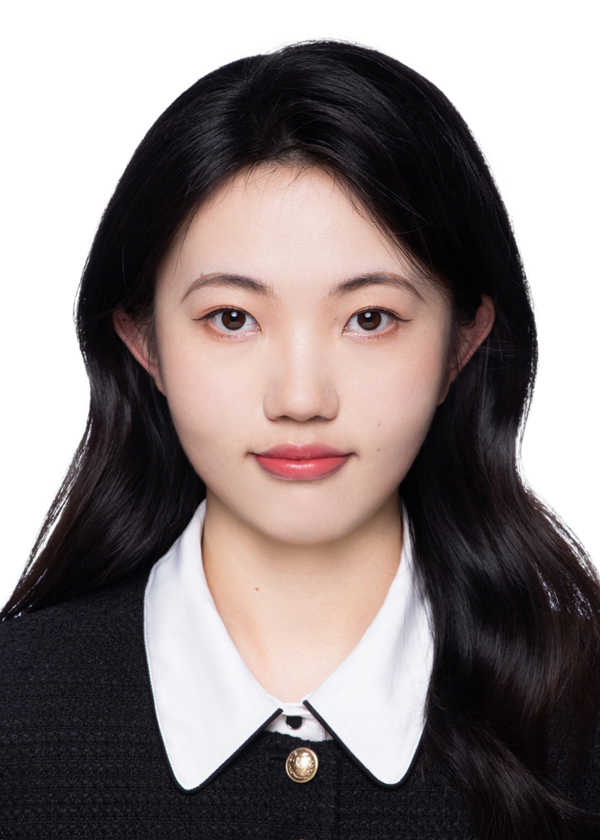}}]
{Jie Lu} 
received the B.Eng. degree in electronics and information engineering from South-Central Minzu University in 2021, and the M.Eng. degree in microelectronics and communication engineering from Chongqing University in 2024. She is currently pursuing the Ph.D. degree with the Department of Computer Science and Engineering, Michigan State University. Her research focuses on AI-driven wireless communications.
\end{IEEEbiography}

\vspace{-0.2in}
\begin{IEEEbiography}
[{\includegraphics[width=1in,height=1.25in,clip,keepaspectratio]{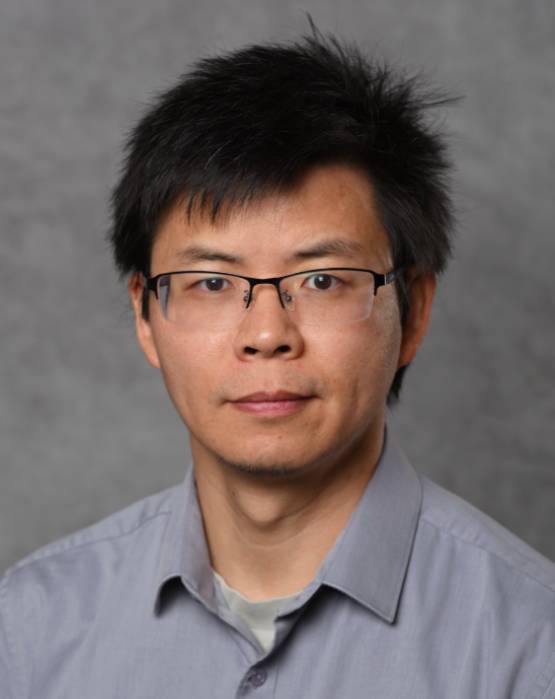}}]
{Huacheng~Zeng} (SM'20) is an Associate Professor in the Department of Computer Science and Engineering at Michigan State University (MSU). Prior to joining MSU, Dr. Zeng was an Assistant Professor in the Department of Electrical and Computer Engineering at the University of Louisville. He also worked as a Senior System Engineer at Marvell Semiconductor, where he contributed to the development of wireless system solutions. He received his Ph.D. in Computer Engineering from Virginia Polytechnic Institute and State University (Virginia Tech). Dr. Zeng is a recipient of the NSF CAREER Award (2019), the Best Paper Award at IEEE SECON (2021), and the Best Student Paper Award at ACM WUWNET (2014). His research interests broadly include wireless networking and mobile sensing systems. He is serving on the editorial board of IEEE Transactions of Wireless Communications.
\end{IEEEbiography}

\vspace{-0.2in}
\begin{IEEEbiography}[{\includegraphics[width=1in,height=1.25in,clip,keepaspectratio]{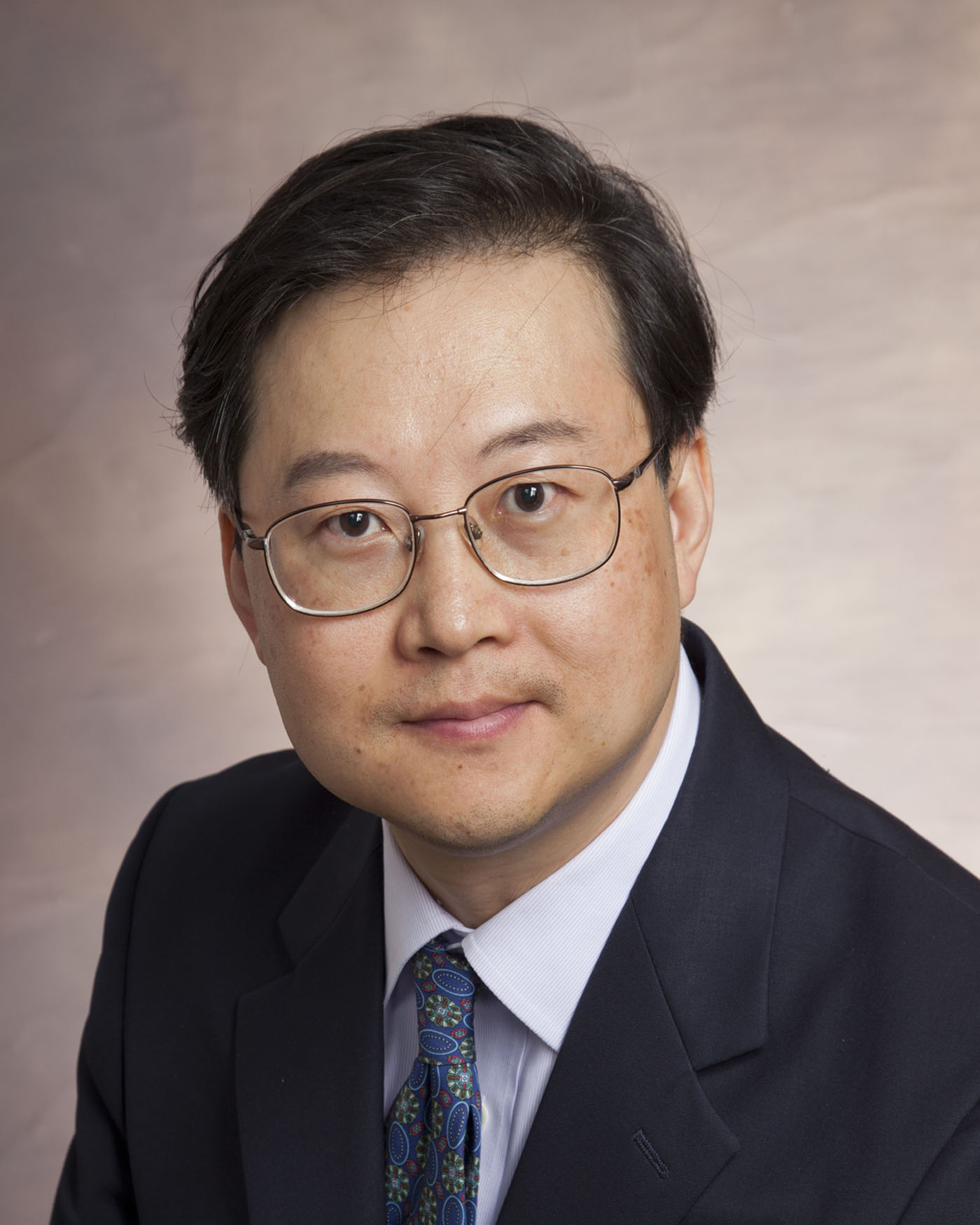}}]{Y. Thomas Hou} 
(Fellow, IEEE)
received his Ph.D. from NYU Tandon School of Engineering in 1998. He is currently Bradley Distinguished Professor of Electrical and Computer Engineering at Virginia Tech, Blacksburg, VA, USA, which he joined in 2002. He was a Member of Research Staff at Fujitsu Laboratories of America in Sunnyvale, CA from 1997 to 2002. 
His current research focuses on developing real-time optimal solutions to complex science and engineering problems arising from wireless and mobile networks. 
He is also interested in wireless security.  
He authored/co-authored two textbooks and has published over 400 papers in IEEE/ACM journals and conferences. 
His papers were recognized by 12 best paper awards from IEEE and ACM, including an IEEE INFOCOM Test of Time Paper Award in 2023. 
He holds six U.S. patents.  
Prof. Hou was named an IEEE Fellow for contributions to modeling and optimization of wireless networks.  
He was/is on the editorial boards of a number of IEEE and ACM transactions and journals. 
He was Steering Committee Chair of IEEE INFOCOM conference and was a member of the IEEE Communications Society Board of Governors.  He was also a Distinguished Lecturer of the IEEE Communications Society.  
\end{IEEEbiography}

\balance

 





\end{document}